# HIGH PRESSURE EFFECTS ON SUPERCONDUCTIVITY

B. Lorenz[1] and C. W. Chu[1,2,3]

[1] Department of Physics and Texas Center for Superconductivity, University of Houston, Houston, Texas 77204-5002, USA
[2] Lawrence Berkeley National Laboratory, 1 Cyclotron Road, Berkeley, CA 94720, USA
[3] Hong Kong University of Science and Technology, Hong Kong, China

## 1. INTRODUCTION

Since the discovery of superconductivity at 4 K in mercury by Kamerlingh Onnes in 1911 the search for higher transition temperatures has been the primary goal for many decades. In the early 1950's B. Matthias discovered a correlation between $T_c$ and the number of valence electrons per atom (R) of a large number of superconducting elements and compounds [1, 2]. While there was no superconducting compound with $R \leq 2$ at the time Matthias showed the existence of a maximum of $T_c(R)$ close to R= 4.75 [1] and a second maximum at R≈ 6.4 [2]. Most of the intermetallic superconductors with a high $T_c$ exhibit lattice instabilities. The application of external pressure to these high $T_c$ superconductors can drive the compounds towards or away from lattice instabilities by varying the principal parameters determining the superconducting properties (the electronic density of states at the Fermi energy, $N(E_F)$, the characteristic phonon frequency, and the coupling constant of electrons and phonons), and it can be used to tune the $T_c$ and the superconducting properties. Almost all of the superconducting metallic elements show a decrease of $T_c$ with pressure. This negative pressure coefficient was attributed to the volume dependence of $N(E_F)$ and of the effective interaction between the electrons mediated by the electron-phonon coupling [3]. However, a few elemental superconductors such as Tl and Re show a more complex pressure dependence of $T_c$ that was explained by pressure-induced changes of the Fermi surface topology [4,5].

Some aspects of the pressure dependence of $T_c$ of low-temperature superconductors (LTS) will be discussed in more detail in Section 2.

New techniques for creating higher pressures have been developed over the last couple of decades pushing the limits of static pressure generation to a few hundred GPa. At high enough pressure almost every ambient pressure structure becomes unstable and transforms into a structure of higher density and, frequently, of higher symmetry. Some elements or compounds undergo a sequence of structural transitions into several different phases under applied pressure. Many of these new structures are metallic and superconductivity was observed in a large number of high-pressure phases even when the compound was not superconducting at ambient pressure. Among the elements of the periodic table, superconductivity induced by pressure has raised the total number of superconducting elements from 29 (at ambient pressure) to 52 including non-metallic elements like sulfur or oxygen and alkali metals with only one valence electron such as Li.

The latter alkali metal is of particular interest since it is one of the elements of the first column of the periodic system that should not be superconducting according to the Matthias rules. In 1969 Allen and Cohen [6], based on their pseudopotential calculations, proposed Li to be superconducting but no evidence of superconductivity has been found at temperatures as low as 4 mK [7]. However, it was predicted that superconductivity should occur in Li under high pressure with transition temperatures of up to 80 K [8]. Early high-pressure experiments [9] revealed a drop of resistivity in Li at 7 K between 22 and 32 GPa but the data were not conclusive enough to be interpreted in terms of superconductivity. Only recent measurements of electrical transport and magnetic properties have unambiguously shown that superconductivity exists in Li at pressures above 20 GPa and at temperatures up to 20 K [10, 11, 12]. The large interest in the superconductivity in dense lithium is also motivated by the fact that it is the "simplest" metal with only one valence electron, no d-electrons in sight, and it is the superconducting element closest to hydrogen. There appears to be new hope in the continuous search for superconductivity in metallic hydrogen that has been predicted to become superconducting at high pressure with an extraordinary high $T_c$ [13].

A new chapter in the ongoing search for new superconducting compounds with higher $T_c$ values was opened with the discovery of superconductivity at $T_c$'s as high as 35 K in the La-Ba-Cu-O (LBCO) cuprate compound in 1986 [14] bypassing the highest known $T_c$ of $Nb_3Ge$ ($T_c$=23.2 K, [15]) by more then 50 %. Soon after this remarkable event Chu et al. [16] reported an unprecedented large positive pressure coefficient of $T_c$ of LBCO that led the authors to conclude that fine-tuning of material parameters by physical and chemical means could lead to even higher transition temperatures. In fact, simulating the effects of external pressure (variation of the lattice constants) by replacing La with the smaller isovalent Y ion ("chemical pressure") led to the next giant leap in raising $T_c$ above the temperature of liquid nitrogen as first demonstrated in the Y-Ba-Cu-O (YBCO) cuprate [17]. A detailed discussion of the pressure effects in this new class of high-temperature superconductors (HTS) is given in Section 3.

With all the success of high-pressure investigations in superconductivity there arises the question what we can learn from these experiments about the microscopic mechanisms, the pairing symmetry, and the validity of theoretical models. High-pressure investigations in conjunction with other experimental data and available theories have been found extremely useful leading to a deeper understanding of the superconducting state and to a possible discrimination between different proposed models. In Section 4 we discuss several recent examples where the results of high-pressure experiments have significantly contributed to a better understanding of the superconductivity in specific compounds or have raised new questions about the mechanisms of superconductivity.

## 2. PRESSURE EFFECTS IN LOW-TEMPERATURE SUPERCONDUCTORS

Soon after the first high-pressure investigations of superconductors had been accomplished it became obvious that the majority of superconducting elements responded to the imposed pressure with a decrease of $T_c$. The fundamental question was raised whether pressure could suppress superconductivity completely if it was just increased to high enough values. The main difficulty at the time was the limited experimentally accessible pressure range of several GPa and the need to extrapolate the experimental data of $T_c(p)$ to much higher pressures to estimate the $T_c=0$ intercept. In many cases $T_c(p)$ turned out to be a non-linear function of p at higher pressures, however, it was proposed [3] that $T_c$ was roughly linear when plotted as function of volume, V. The linear $T_c(V)$ relationship was later shown to hold for a large number of superconducting metals and the critical pressures above which superconductivity cannot persist have been derived from a linear extrapolation of $T_c(V)$ [18]. An impressive example of this linear $T_c$ vs. V relation is shown in Fig. 1. Based on the extrapolation of the results shown in Fig. 1 and using the known compressibility data the critical pressures for Al, Cd, and Zn were estimated as 6.7 GPa, 3.8 GPa, and 4.1 GPa, respectively [18].

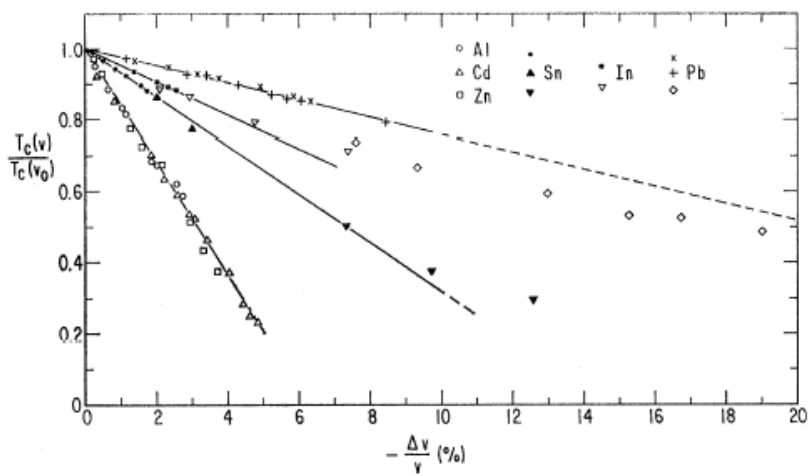

**Fig. 1**: The relative pressure shift of $T_c$ as function of the relative volume change for a large number of superconducting metallic elements (Figure reproduced from Ref. [18]).

Contrary to the suppression of superconductivity by reducing the volume (or lattice constants), an increase of $T_c$ could be expected by expanding the lattice. This possibility that was already discussed in the 1950's eventually led B. Matthias to enlarge the lattice parameters in a solid solution of NbC and NbN and to set a new record for $T_c$ (17.8 K) at this time [1].

## 2.1 The Pressure Dependence of $T_c$ in BCS and Strong-Coupling Theory of Superconductivity

The pressure dependence of $T_c$ can be derived from microscopic theories of phonon-mediated superconductivity such as the weak-coupling BCS model [19] or the Eliashberg theory of strong-coupling superconductivity [20]. For a specific superconductor the knowledge about the pressure dependence of its microscopic parameters such as the average phonon frequency, $\omega_D$, the electronic density of states, $N(E_F)$, etc. is than needed to estimate $T_c$ as function of imposed pressure.

According to the weak-coupling BCS theory $T_c$ is expressed by [19]

$$k_B T_c = 1.13 \hbar \omega_D \exp\{-1/\lambda\} \quad (1)$$

$k_B$ and $\hbar$ are Boltzman's and Planck's constant, respectively, $\lambda = N(E_F)V_{eff}$, and $V_{eff}$ is the effective interaction between the electrons mediated by the electron-phonon coupling. Since $\omega_D$ commonly increases with pressure (phonon hardening) the frequently observed decrease of $T_c$ with p must be due to a decrease of $\lambda$. The average value of the electronic density of states is expected to decrease under pressure because of the pressure-induced band broadening effect. However, the pressure dependence of $N(E)$ can be very different at the Fermi energy $E_F$, depending on the details and the topology of the Fermi surface in a particular compound. The pressure dependence of $V_{eff}$ is even more difficult to estimate and it needs a microscopic treatment of the coupled system of electrons and phonons. It should be noted that the BCS-equation (1) applies only in the weak-coupling case when the electron-phonon coupling constant $\lambda$ is small.

Most superconducting elemental metals (except Al) exhibit an intermediate to strong electron-phonon coupling with $\lambda$ of the order of 1 or larger [21] and their superconductivity is not well described by the weak-coupling BCS equation (1). Extending the original BCS theory [19], Eliashberg developed the theory of strong-coupling superconductivity [20] that provides, together with the pseudopotential treatment of the screened Coulomb interaction [22], the fundamental equations describing the physics of superconductors with phonon mediated pairing of any coupling strength. By solving the finite-temperature Eliashberg equations McMillan [23] derived an equation for the transition temperature valid in the strong coupling case that was later slightly modified by Dynes [24]:

$$k_B T_c = \frac{\hbar \langle \omega \rangle}{1.2} \exp\left\{-\frac{1.04(1+\lambda)}{\lambda - \mu^*(1+0.62\lambda)}\right\} \quad (2)$$

$\mu^*$ is the Coulomb pseudopotential calculated in [23]:

$$\mu^* = \frac{N(E_F)V_c}{1 + N(E_F)V_c \ln \frac{E_B}{\hbar \omega_0}} \quad . \quad (3)$$

$V_c$ is the matrix element of the screened Coulomb interaction averaged over the Fermi surface, $E_B$ is the electronic bandwidth and $\omega_0$ is the cut-off (or maximum) phonon frequency. The electron-phonon coupling constant $\lambda$ is given by [23]

$$\lambda = 2\int \frac{d\omega \alpha^2(\omega) F(\omega)}{\omega} = \frac{N(E_F)\langle I^2 \rangle}{M\langle \omega^2 \rangle} \quad . \tag{4}$$

$\alpha(\omega)$ and $F(\omega)$ are the strength of an average electron-phonon interaction and the phonon density of states, respectively. $\langle I^2 \rangle$ is the average over the Fermi surface of the square of the electronic matrix element of the change of the crystal potential induced by a displacement of an atom. M stands for the atomic mass. $\langle \omega^n \rangle$ in equations (2), n=1, and (4), n=2, is the n-th moment of the normalized weight function:

$$\langle \omega^n \rangle = \frac{2}{\lambda}\int d\omega\, \alpha^2(\omega)\, F(\omega)\omega^{n-1}$$

From equations (2) to (4) it becomes obvious that the pressure dependence of $T_c$ is very complex and depends on various parameters of the electron and phonon systems. The pressure effects on $\mu^*$ and on $\langle I^2 \rangle$ are frequently neglected which simplifies the evaluation and reduces the pressure dependence to two parameters, the density of states $N(E_F)$ and the average phonon energy $\langle \omega \rangle$ (or $\langle \omega^2 \rangle$). The McMillen formula (2) has been used successfully in a number of examples in deriving $T_c(p)$ from first principle calculations of the pressure effect on the band structure and the phonon spectrum. For example, Olsen et al. [25] have demonstrated that equation (2) can be used to estimate the pressure dependence of $T_c$ for a number of non-transition-metal superconductors in qualitative agreement with experiments.

The pressure dependence of the superconducting $T_c$ of lead was investigated very thoroughly, experimentally [18, 26] as well as theoretically [27]. For the high-pressure experimentalist the use of lead as an in-situ manometer at temperatures below 7 K appeared to be very attractive. Since lead is easily available in high purity (99.9999 %), its superconducting transition is extremely sharp, and the initial pressure coefficient, $dT_c/dp = -0.386\, K/GPa$, is of a reasonable magnitude it has advanced to a secondary pressure gauge that allows the accurate pressure measurement at low temperature in various pressure cells. A detailed theoretical treatment of the pressure effects on $T_c$ of lead was presented by Hodder [27]. Neglecting the pressure dependence of $\mu^*$ in equation (2) and assuming a multiple-cutoff Lorentzian model [28] for the phonon density of states $F(\omega)$ the pressure dependence of the electron-phonon coupling constant could be related to the pressure dependence of the longitudinal ($\omega_l$) and transverse ($\omega_t$) phonon frequencies. The pressure effect on the electronic matrix element was calculated from a pseudopotential theory and found to be small compared with the pressure shift of the phonon energies. With the known values of $d\ln\omega_l/dp$ and $d\ln\omega_t/dp$ the pressure coefficient of the $T_c$ of lead was estimated as $d\ln T_c/dp = -0.05\, GPa^{-1}$. This value is in very good agreement with the known experimental data [18].

## 2.2 The Non-Linear $T_c(p)$ of Thallium and Rhenium - Pressure Revealing Electronic Instabilities of the Fermi Surface

The majority of the superconducting elements show a rather monotonous decrease of $T_c$ with pressure and a linear $T_c(V)$ as shown in Fig. 1. However, there are a few exceptions, thallium and rhenium, that exhibit a very unusual $T_c(p)$. The critical temperature of thallium was found to increase at low pressures. It passes through a maximum at about 0.2 GPa and decreases at higher p [3,4]. Since no structural transition was observed under pressure it was suspected that the unusual $T_c(p)$ dependence might be due to a modification of the electronic configuration under pressure [4]. Lazarev et al. [29] proposed a change in the Fermi surface topology as the possible explanation for the anomalous pressure dependence of $T_c$ of Tl. A non-linear contribution to the pressure dependence of $T_c$ was indeed shown [30] to result from a change of the Fermi surface topology (Lifshitz instability, [31]) from a closed surface to an open one by forming of a neck. This topological transition causes changes of the density of states, thermodynamic parameters, as well as the superconducting gap.

A similar anomalous change of $T_c$, but in opposite direction, with pressure was also observed in rhenium [5]. $T_c$ of Re first decreases at low pressure, passes through a minimum close to 0.7 GPa, and increases slightly above this pressure. $T_c(p)$ measured for a Re single-crystal [32] is shown in Fig. 2.

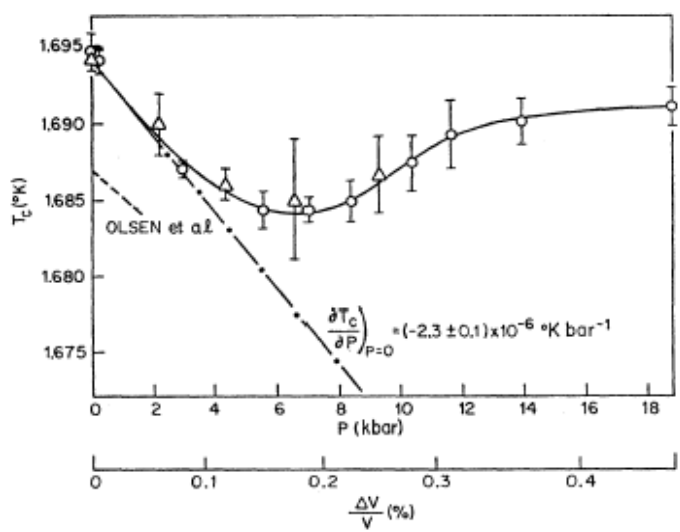

**Fig. 2**: The pressure shift of $T_c$ of a single crystal of rhenium (Figure reproduced from Ref. [32]).

Based on the $T_c(p)$ data shown in Fig. 2 and the similarity to the thallium data Chu et al. proposed to adopt the approach of Ref. [29] and to express the pressure dependence of $T_c$ of rhenium by the superposition of a linear variation of $T_c$ with p and a nonlinear contribution due to a pressure-induced change of the Fermi surface topology. The model is supported by the observation that alloying rhenium with a minor amount of other metals (like osmium, tungsten, or molybdenum) has a dramatic effect on the anomalous $T_c(p)$ dependence [5,32]. For example, for Re-Os alloys the minimum in the $T_c(p)$ curves shifts to lower pressure with increasing Os concentration and, eventually, disappears at a critical concentration of about 0.2

%. The $T_c(p)$ data for the Re-Os alloys are displayed in Fig. 3. At Os concentrations in excess of 2.75 % the pressure dependence of $T_c$ is close to linear and can be considered as "normal". A similarly strong effect was observed in Re-W alloys but, in contrast to the Re-Os alloy, the critical pressure defined by the minimum of $T_c(p)$ increased quickly with the tungsten concentration [32]. This qualitatively different behavior can be understood by taking in mind that Os adds one electron thus raising the Fermi energy whereas W decreases the number of electrons and lowers $E_F$. Both contributions to the pressure dependence of $T_c$, the "normal" linear decrease of $T_c$ with p and the nonlinear part (due to the change of Fermi surface topology) could be separated and it was found that in pure Re the nonlinear contribution starts at pressures above 0.2 GPa. A more detailed discussion of the topological changes of the Fermi surface that will appear if the Fermi energy passes through critical points in the reciprocal space and its relation to the nonlinear component of $T_c(p)$ is given in Ref. [32]. It should be noted that a band structure calculation for rhenium [33] revealed the existence of such critical points near the Fermi surface and provided evidence for the change of Fermi surface topology similar to that discussed above when the Fermi Energy is raised.

The anomalous pressure shifts of $T_c$ observed in thallium and rhenium reveal the important influence of the electronic structure, the density of states, and the Fermi surface topology on the superconducting properties of LTS.

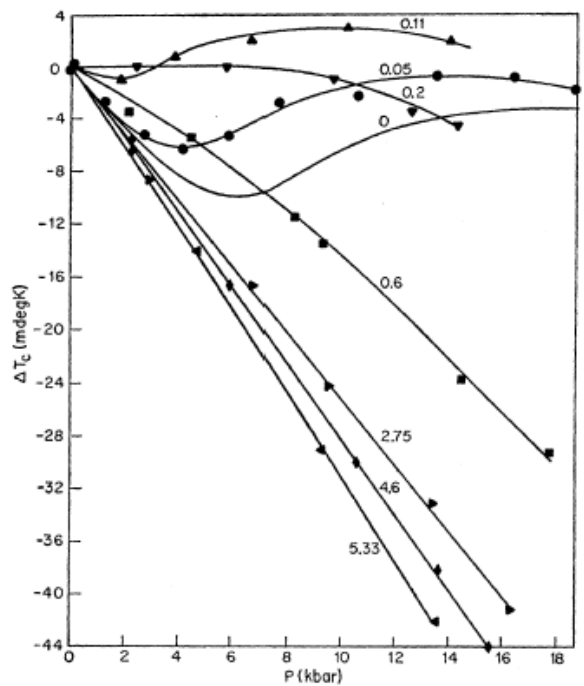

**Fig. 3**: The pressure shift of $T_c$ for Re-Os alloys. The Os concentration in at. % is indicated by the numbers next to the curves. (Figure reproduced from Ref. [32]).

## 2.3 Pressure Dependence of $T_c$ and Lattice Instabilities – The Common Electronic Origin of Superconductivity and Structural Transformations

The A15 intermetallic compounds $A_3B$ have attracted attention because of the high transition temperatures shown (up to 23.2 K for $Nb_3Ge$). Some A15 compounds, such as $V_3Si$, are weakly coupled BCS-like superconductors but others (e.g. $Nb_3Sn$, $Nb_3Ge$) exhibit strong electron-phonon coupling and are better described by the strong-coupling theory. A number of A15 compounds undergo a structural phase transition at $T_L > T_c$ from the high-temperature cubic phase to a low-temperature tetragonal phase (although the tetragonal distortion is very small). A soft phonon mode has been observed in almost all A15 intermetallics with high $T_c$ and it is the driving force for the cubic-to-tetragonal distortion at $T_L$ observed in some compounds. Therefore, it was speculated that the soft phonon modes play a major role in stabilizing superconductivity at high critical temperatures.

To investigate the interplay between electronic excitations and soft phonon modes leading to structural instabilities and its significance for stabilizing superconductivity with high $T_c$'s it is desirable to tune the structural transition temperature closer to the superconducting $T_c$ by either chemical or physical means. Since the A15 compounds $A_{3+x}B_{1-x}$ have very narrow ranges x of homogeneity it is more difficult to change $T_L$ and $T_c$ by tuning the chemical composition. The application of pressure, however, was shown to be a powerful tool to control the cubic-to-tetragonal transition in several A15 compounds [34,35].

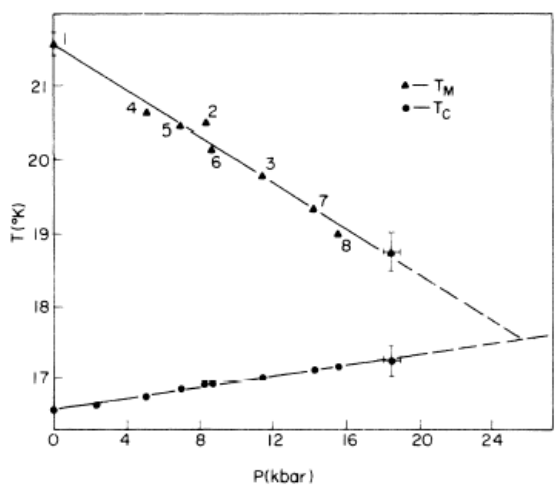

**Fig. 4**: The pressure shifts of $T_L$ (triangles) and $T_c$ (circles) in $V_3Si$. (Figure reproduced from Ref. [34]).

The pressure dependences of $T_L(p)$ and $T_c(p)$ were simultaneously measured in $V_3Si$ single crystals [34] and a strong correlation was found between $T_L$ and $T_c$ (Fig. 4). Whereas $T_L$ decreases linearly with pressure the superconducting $T_c$ increased towards $T_L$. At the critical pressure of $p_c \approx 2.4$ GPa (estimated by extrapolation) both transitions happen at the same temperature, $T_L(p_c)=T_c(p_c)$, i.e. the lattice transformation is suppressed down to the superconducting state. The structural transition was theoretically described within the Weger-Labbé-Friedel (WLF) linear-chain model [36] and it was shown that good agreement with the experimentally derived pressure coefficient $dT_L/dp$ was only obtained for $V_3Si$ when the interband charge transfer under pressure was properly taken into account [37]. The calculation shows that the pressure effect on the electronic structure and the pressure-induced charge transfer from s to d-bands are essential to explain the experimental data. However, the experimental results shown in Fig. 4 up to 1.8 GPa are also consistent with the suggestion that $T_c$ in $V_3Si$ could be enhanced by the softness of the lattice induced by pressure [38]. Within this model $T_c(p)$ is expected to pass through a maximum at $p_c$ ($T_c=T_L$) where the lattice softness is also at its maximum [38]. To distinguish between the proposed mechanisms for the lattice instability and the $T_c$-enhancement under pressure it appeared necessary to measure $T_c(p)$ at and beyond the critical pressure $p_c$.

$V_3Si$ is a special A15 compound in that the cubic-to-tetragonal transition can be suppressed depending on the sample preparation and conditions. The nontransforming (NT) samples show a weaker phonon softening effect and their $T_c$'s are found only slightly higher than those of transforming samples. The pressure coefficient of $T_c$ of the NT $V_3Si$ single crystal, however, was shown to be positive and lower (by about 30 %) than $dT_c/dp$ of the $V_3Si$ single crystal that exhibits the structural change from cubic to tetragonal symmetry [39]. This result suggests that the $T_c$ of a transforming sample should still increase with pressure if p exceeds the critical value $p_c$ above which the structural instability is completely suppressed. Extending the experimental pressure range to 2.9 GPa Chu and Diatschenko could later confirm this prediction [40]. By comparing the pressure coefficients of $T_c$ of transforming as well as NT $V_3Si$ single crystals it was unambiguously shown that $T_c(p)$ of the transforming sample does not exhibit the maximum predicted by the soft-phonon model but it changes slope above $p_c$ and the pressure coefficient for $p>p_c$ was found to be very close to the value of the NT sample. Close to the highest pressure of this experiment a sudden drop of $T_c$ of the NT sample indicated a pressure-induced phase transition, possibly into a non-superconducting phase, as was previously inferred from the negative pressure coefficient of the shear modulus [41].

In the WLF model $T_L$ and $T_c$ are non-monotonous functions of the number of d-electrons Q in the sub-band. The pressure-induced redistribution of charges between different bands will change Q and the position of the Fermi level. At ambient pressure $E_F$ is close to a peak of the density of states and the application of pressure can enhance or suppress the lattice instability ($T_L$) if $E_F$ is moved either towards or away from the peak, respectively. The analytic dependence of the two critical temperatures $T_L(Q)$ and $T_c(Q)$ have been calculated within the WLF model for $V_3Si$ and $Nb_3Sn$ [42]. It was shown that the functions $T_L(Q)$ and $T_s(Q)$ exhibit a maximum at $Q_L$ and $Q_c$, respectively, and $Q_L<Q_c$ for $V_3Si$ and $Q_c<Q_L$ for $Nb_3Sn$. For values of Q in between $Q_L$ and $Q_c$ the changes of $T_L(Q)$ and $T_c(Q)$ are opposite in sign [42]. Since pressure increases Q (charge transfer to the d-band) the p-induced increase of $T_L$ and decrease of $T_c$ observed in $V_3Si$ is qualitatively understood ($Q_L<Q<Q_c$). A quantitative analysis is given in Ref. [39]. In $Nb_3Sn$, however, the charge transfer must result in the opposite changes of $T_L$ (increasing with Q) and $T_c$ (decreasing with Q) as long as $Q_c<Q<Q_L$. The enhancement of the lattice instability in $Nb_3Sn$ and the decrease of $T_c$ with pressure were in fact observed (see Fig. 5) and a qualitative interpretation of the pressure effects based on the WLF model was given [35]. The success of the WLF model with the inclusion of the pressure-induced interband charge transfer in describing the pressure dependence of $T_L(p)$ and $T_c(p)$ in a number of A15 compounds unambiguously demonstrates the superior role of the electronic excitation spectrum with peaks or singularities of the density of states close to the

Fermi energy for inducing the phonon softness and lattice instabilities as well as superconductivity with high $T_c$-values. The high-pressure experiments in A15 compounds discussed in the last paragraphs have shed new light onto the complex problem of structural instabilities, superconductivity, and the electronic structure. Similar phenomena are to be expected in other materials where superconductivity and structural instabilities exist in close neighborhood.

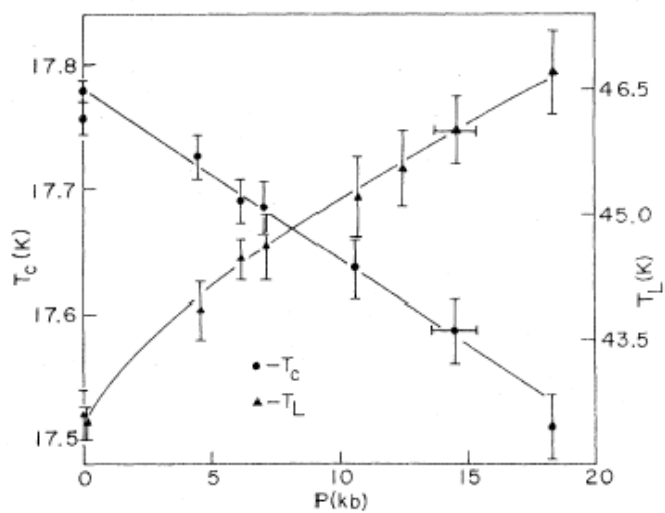

**Fig. 5**: The pressure shifts of $T_L$ (triangles) $T_c$ (circles) in $Nb_3Sn$. (Figure reproduced from Ref. [35]).

The vanadium-ruthenium alloys close to the equiatomic composition undergo a cubic to tetragonal transition upon cooling. $T_L$ decreases with decreasing Ru concentration and the transition disappears below a critical value of $C_{crit} \approx 45.5$ at.% of Ru. Just above $C_{crit}$ superconductivity was shown to arise with a sharp peak of $T_c$ right at the critical concentration [43]. From ambient pressure measurements of transport and magnetic properties of a series of V-Ru alloys with compositions around $C_{crit}$ it was suggested that the origin of the structural transition into the tetragonal phase lies in an electronic instability arising from a substantial reduction of the Fermi surface area and the associated decrease of the electronic contribution to the free energy in the tetragonal structure. Alternatively, the soft phonon model based upon the maximum of the lattice softness at $C_{crit}$ and its effect on the superconducting temperature was proposed to explain the sharp maximum of $T_c$ [44]. High-pressure experiments have been conducted to discriminate between the different proposals [45]. Fig. 6 (left graph) shows that the tetragonal phase of $V_{0.54}Ru_{0.46}$ is completely suppressed at a critical pressure of $p_c \approx 1.4$ GPa. The pressure dependence of $T_c$ is displayed in Fig. 6, right graph. At low pressures $T_c$ increases linearly with p, it changes slope at $p_c$ and it increases further at a reduced rate for $p>p_c$. This observation cannot be explained by the lattice softening mechanism that predicted a distinct maximum of $T_c(p)$ right at $p_c$ but it lends strong support to the model of electronic instability [44,45].

The systematic investigation of the pressure effects on the superconducting $T_c$ and the structural instabilities in various compounds have led to the proposition that the electronic spectrum plays the major role in both, superconducting and lattice transformations. The suggestion is also supported by the correlation between the structural instability and superconductivity as revealed by high-pressure experiments in layered charge-density-wave compounds, for example 2H-NbSe$_2$ [46].

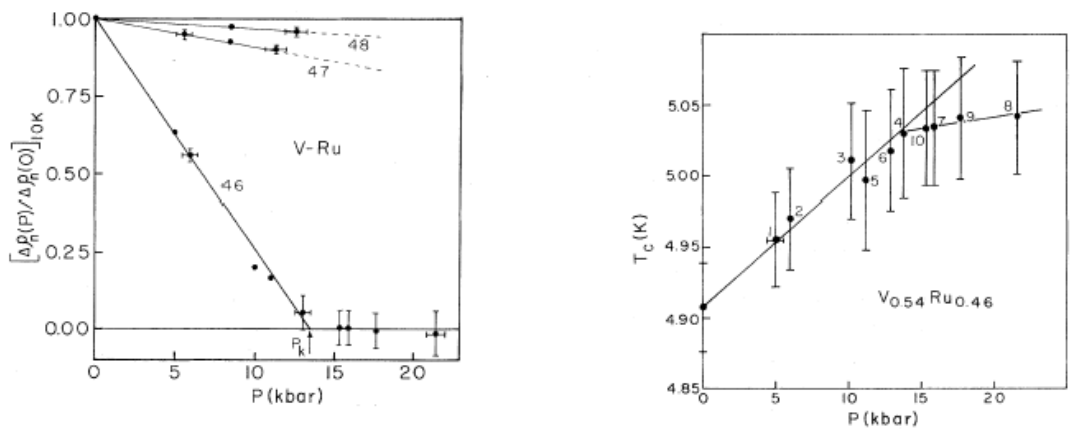

**Fig. 6**: Suppression of the tetragonal phase in V-Ru alloys with pressure (left) and the pressure dependence of $T_c$ for $V_{0.54}Ru_{0.46}$ (right). (Figure reproduced from Ref. [45]).

## 3. PRESSURE EFFECTS IN HIGH-TEMPERATURE SUPERCONDUCTORS

The discovery of superconductivity in the La-Ba-Cu-O compound system [14] with unprecedented high transition temperatures of 35 K has revived a tremendous activity in the field of superconductivity. The positive pressure coefficient frequently observed in these new compounds have directed scientists to synthesize new materials with even higher $T_c$'s. In the following Sections we focus our discussion on a few topics of high-pressure investigations of HTS.

### 3.1 The Increase of $T_c$ to New Records Under Imposed Pressure

The very first experiment on the pressure shift of $T_c$ of the new La-Ba-Cu-O (LBCO) compound revealed a surprising result. The superconducting transition temperature was raised above 40 K with pressure at an unusually large rate [16]. The $T_c(p)$ data of Ref. [16] are shown in Fig. 7. The enormous increase of $T_c(p)$ at pressures up to 1.3 GPa (the data at higher pressure are invalid since the sample was destroyed under pressure) and the missing signs of saturation of $T_c$ immediately led to the conclusion that even higher $T_c$ values should be achievable at higher pressures and for optimized samples. Furthermore, it became obvious that the superconducting phase in the LBCO system is the one with the K$_2$NiF$_4$ structure, i.e. La$_{2-x}$Ba$_x$CuO$_{4-\delta}$ [16]. With this knowledge in mind C. W. Chu et al. [47] succeeded to optimize the sample preparation and to synthesize compounds with the nominal composition La$_{1.8}$Ba$_{0.2}$CuO$_{4-\delta}$. The onset of superconductivity in these compounds increased under hydrostatic pressure to 52.5 K [47], a new record $T_c$ in early 1987.

There are two remarkable results from the early investigations of the new class of superconducting compounds: (i) The superconducting phase was identified as the perovskite ($K_2NiF_4$) structure. (ii) Pressure increases $T_c$ at an exceptionally high rate (more than 6 K/GPa in the example shown in Fig. 7). This is in contrast to many low-$T_c$ compounds showing a negative pressure coefficient of $T_c$. The LTS with an increasing $T_c(p)$ (such as $V_3Si$, Section 2.3) exhibit a much smaller coefficient $dT_c/dp$ than that observed in the La-Ba-Cu-O system.

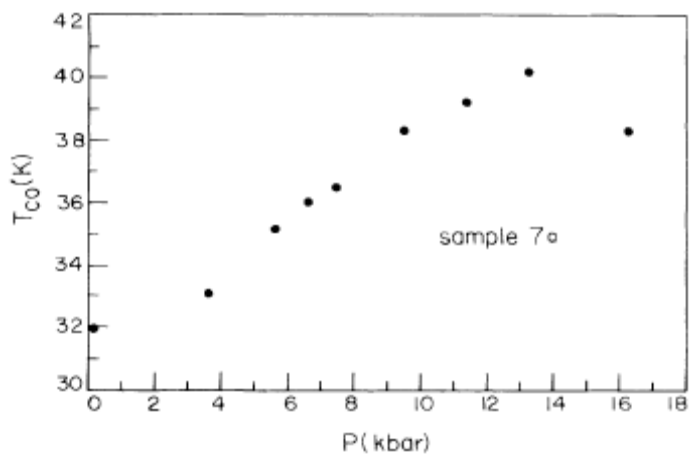

**Fig. 7**: Pressure dependence $T_c(p)$ of the HTS superconductor $La_{2-x}Ba_xCuO_{4-y}$. (Figure reproduced from Ref. [16]).

The large positive effect of external pressure on $T_c$ of LBCO led to the next important step in the search for new superconducting compounds that eventually raised the transition temperature above the temperature of liquid nitrogen. In simulating the external pressure effect (lattice contraction) by replacing $La^{3+}$ with the smaller ion $Y^{3+}$ a new superconducting compound system, Y-Ba-Cu-O (YBCO), was synthesized with an onset temperature of $T_c$=93 K [17]. Interestingly, the application of external pressure did not change the $T_c$ notably but rather caused a broadening of the resistive transition into the superconducting state [48]. The origin of this striking difference to the large pressure coefficient of LBCO was attributed to the presence of the large chemical pressure associated with the smaller $Y^{3+}$ ions. The superconducting phase in the YBCO system was later identified as the tetragonal $YBa_2Cu_3O_{6+\delta}$ and it was shown that a homologous series of isostructural compounds can be derived by replacing Y with different rare earth ions, such as La, Nd, Sm, Eu, Gd, Ho, Er, and Lu, with superconducting $T_c$'s all above 90 K [49]. These results are particularly interesting because they prove that not the size of the A-ion in the structure of $A-Ba_2Cu_3O_{6+\delta}$ determines the high $T_c$ but it is the layered structure with $CuO_2$-Ba-$CuO_{2+\delta}$-Ba-$CuO_2$ slabs separated by the A-ions and stacked along the c-axis that gives rise to the large $T_c$ values.

In the following years other HTS compounds with even higher $T_c$'s have been synthesized with an increasing structural complexity (for a systematic review of HTS structures see Ref. [50]). The replacement of the (relatively expensive) rare earth ions by non-group-IIIb elements led to the discovery of the Tl-based superconducting compounds Tl-Ba-Cu-O ($T_c$≈90 K, [51]) and Tl-Ca-Ba-Cu-O (TCBCO, $T_c$>100 K, [52]) as well as the Bi-

Al-Ca-Sr-Cu-O compound system (onset-$T_c$=114 K, [53]). The effect of hydrostatic pressure on $T_c$ was investigated for the Bi-HTS compound. Initially, $T_c$ increases with pressure at a rate of 3 K/GPa, it passes through a maximum and decreases above 1.2 GPa [53]. This was the first time that a maximum of $T_c(p)$ was observed in an HTS compound.

The TCBCO high-temperature superconductors can be synthesized in different layered structures described by the general formula $Tl_2Ca_{n-1}Ba_2Cu_nO_{2n+4-\delta}$ with e.g. n=1 (Tl-2021), n=2 (Tl-2122), and n=3 (Tl-2223). The ambient-pressure $T_c$ of this series varies between 0 and 125 K. Tl-2021 is of particular interest since it is an overdoped HTS and $T_c$ can be changed between 0 and 90 K by solely varying the oxygen content. The pressure dependence of $T_c$ of TCBCO (n=1,2,3) was investigated by Lin et al. [54]. For Tl-2122 and Tl-2223 the $T_c$ was found to increase at a moderate rate of 1.8 K/GPa and 2.4 K/GPa, respectively, in accordance with all previous high-pressure data of HTS. However, the pressure coefficients of $T_c$ of Tl-2021 samples at different doping levels were all shown to be negative with $dT_c/dp$ between –3.9 K/GPa and –1.4 K/GPa, depending on the doping state (or the oxygen content). The large negative values of $dT_c/dp$ for Tl-2021 were also reported almost simultaneously by Môri et al. [55] and Sieburger et al. [56]. The significance of $dT_c/dp<0$ of a hole-doped HTS compound for the understanding of HTS was pointed out in Ref. [54]. In particular, the proposed asymmetry between the pressure dependence of $T_c$ of hole-doped and electron doped HTS compounds [57] seemed not to be valid. Based on all available data for $dT_c/dp$ of HTS materials a correlation between $dlnT_c/dp$ and $T_c$, shown in Fig. 8, was shown to exist [50,54]. From this correlation it was proposed that there exists one pressure-sensitive parameter that determines $T_c$ of all hole-doped HTS, such as the carrier concentration n or $n/m^*$ ($m^*$ effective carrier mass), as also suggested from Hall effect [58] and penetration depth measurements [59]. The carrier number dependence of $T_c$ and its effect on the pressure coefficient of $T_c$ based on a charge transfer model will be discussed in Section 3.2.

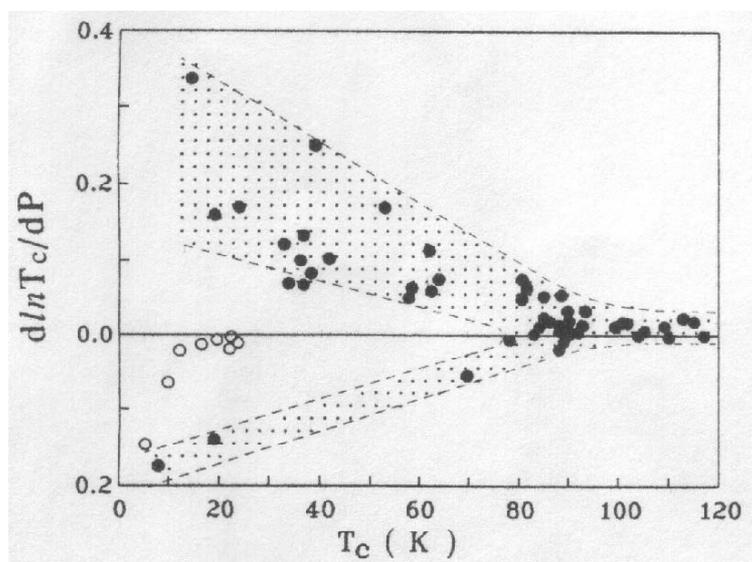

**Fig. 8**: $dlnT_c/dp$ vs. $T_c$ for a large number of HTS compounds. The data for hole-doped HTS (closed circles) can be grouped into two bands joining at the highest $T_c$'s close to $dlnT_c/dp$=0. Open circles: data for electron-doped HTS. (Figure reproduced from Ref. [50]).

The continuous enhancement of $T_c$ with p of Tl-2223 observed at pressures below 2 GPa [54,60] suggested that even higher $T_c$ values could be achieved in this compound if the pressure was further increased. For a slightly under-doped single crystal of $Tl_2Ca_2Ba_2Cu_3O_{10-\delta}$ Berkley et al. [61] reported an increase of $T_c$ from 116 K (p=0) to 131.8 K at 7.5 GPa. This was a new record $T_c$ at this time.

The search for new HTS materials with even higher $T_c$'s culminated in 1993 with the discovery of the mercury based HTS compounds. After the successful synthesis of the one-layer compound $HgBa_2CuO_{4+\delta}$ (Hg-1201) with $T_c$=94 K [62] Schilling et al. [63] reported superconductivity below 133.5 K in the three-layer system $HgBa_2Ca_2Cu_3O_{8+\delta}$ (Hg-1223). This remarkable result was immediately confirmed by other groups [64] with onset transition temperatures of the resistivity drop as high as 140 K. Until today no other compound has been found with higher superconducting temperature than Hg-1223. The positive pressure coefficient of $T_c$ ($dT_c/dp \approx 1.8$ K/GPa, [64]) indicated that superconductivity at even higher temperatures could be stabilized under high-pressure conditions. Upon further increase of the imposed pressure the $T_c$ of Hg-1223 (as derived from the onset of the resistive transition) was shown to increase to 153 K at 15 GPa [65], 157 K at 23.5 GPa [66], and 150 K at 11 GPa [67], respectively. No maximum or saturation of $T_c(p)$ was detected at the highest pressures of 23.5 GPa although the pressure coefficient appeared to decrease with increasing p. Far higher pressures were needed to get to the largest possible $T_c$ in the Hg-1223 compound. Extending the experimental pressure range to 45 GPa Gao et al. [68] achieved the highest ever reported superconducting transition temperature of 164 K in Hg-1223. The data for $T_c(p)$ for the three Hg-based HTS structures, Hg-1201, Hg-1212, and Hg-1223, are shown in Fig. 9.

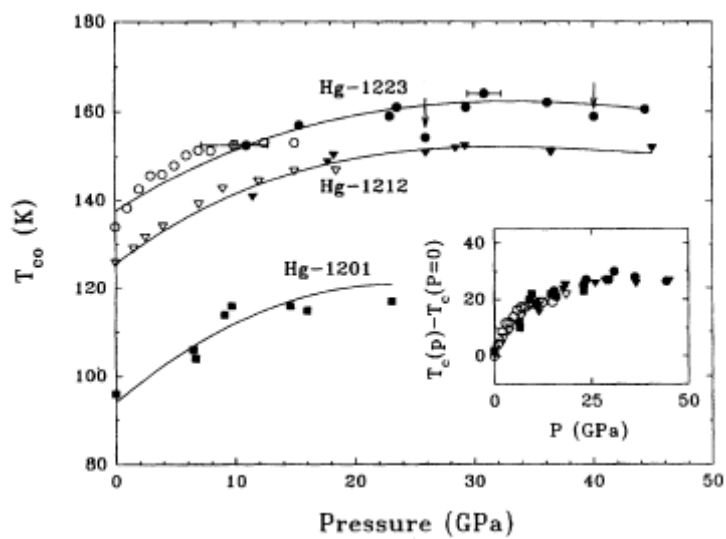

**Fig. 9**: Pressure dependence of $T_c$ for Hg-1201, Hg-1212, and Hg-1223. Open symbols: data from previous reports, solid symbols: data from the diamond anvil cell experiment. (Figure reproduced from Ref. [68]).

There is clear evidence from the $T_c(p)$ data of Fig. 9 that a maximum of $T_c$ exists close to 30 GPa. At higher pressures $T_c$ finally decreases again. This tendency is characteristic for HTS compounds with an initial positive $dT_c/dp$ [50,69]. However, the magnitude of the $T_c$ enhancement, $\Delta T_c(p)=T_c(p)-T_c(0)$, of up to 30 K is unusually large. Interestingly, $\Delta T_c(p)$ is nearly independent of the number of $CuO_2$ layers, as shown in the inset of Fig. 9 where $\Delta T_c(p)$ is plotted for Hg-1201, Hg-1212, and Hg-1223. The large $\Delta T_c(p)$ could be related to the enhanced compressibility (as compared to other HTS materials) of the Hg-HTS compounds [70]. Despite intense research to further increase the superconducting critical temperature of HTS by chemical or physical means the 164 K $T_c$ of Hg-1223 under pressure is still, after more than one decade, the accepted record value for all known superconducting compounds. Whether higher $T_c$ values can be achieved in the known HTS structures or in completely new compounds in the near future is uncertain and any projection has to be speculative.

### 3.2 The Unified Phase Diagram, the Charge Transfer Model, and the Understanding of Pressure Effects in HTS

Many attempts have been undertaken to understand the pressure effects in HTS materials. The key lies in the highly anisotropic layered structure of the HTS compounds, where superconductivity arises in nearly 2-dimensional $CuO_2$ layers. With a few exceptions (like the LBCO-214 structures of HTS) the active $CuO_2$ plane is next to a layer of e.g. SrO or BaO including an apical oxygen ion close to the $CuO_2$ plane. This block is considered a "charge reservoir" since charges are transferred to the $CuO_2$ planes via the apical oxygen resulting in an intermediate valence of the Cu in the active layer. Additional layers are inserted into the structure forming a wealth of HTS compounds [50]. The superconducting HTS compounds are derived from insulating parent structures by introducing hole carriers into the $CuO_2$ layers. The hole density n (or the ratio $n/m^*$) was shown to be one of the crucial parameters of HTS [50,58,59]. An empirical formula correlating the $T_c$ of HTS with the hole density in the $CuO_2$ planes was proposed by Tallon and shown to hold for a large number of HTS compounds [71]:

$$T_c/T_{c,\max} = 1 - a(n-n_{opt})^2 \ , \quad n_{opt} \approx 0.16 \ , \quad a = 82.6 \ . \tag{5}$$

According to the parabolic dependence (5) $T_c(n)$ exhibits a maximum at the optimal hole density in the $CuO_2$ planes, $n_{opt}$. Depending on the doping level the HTS compounds are classified as under-doped, optimally doped, and over-doped for $n<n_{opt}$, $n=n_{opt}$, and $n>n_{opt}$, respectively. The $T_c(n)$ dependence adds an additional degree of freedom to the pressure effect on $T_c$ of HTS. Whereas most hole-doped HTS compounds show a positive pressure coefficient of $T_c$ it was demonstrated that compounds with the highest $T_c$ values (i.e. close to optimal doping) exhibit a very small pressure effect only and in some over-doped cuprates (e.g. $Tl_2Ba_2CuO_{6-\delta}$) $T_c$ actually decreases with pressure. A universal trend correlating $d\ln T_c/dp$ with $T_c$ was proposed and the different magnitudes and signs of $d\ln T_c/dp$ for under-doped, optimally doped, and over-doped HTS were attributed to a pressure-induced increase of the hole density due to a charge transfer from the reservoir to the $CuO_2$ layers [54]. Further experimental evidence from positron lifetime investigations [50] and Hall effect measurements [72] under pressure provides additional support of the charge transfer model. It is the strength of the model that, based on the known parabolic dependence $T_c(n)$, it explains naturally the crossover of dlnT$_c$/dp with increasing degree of doping from positive values in the under-doped region to negative numbers in the over-doped range. The carrier density n increases roughly linear with pressure and is of the order of 8 to 10%/GPa [50, 72]. Since the universal parabolic T$_c$(n) relation is symmetric around n$_{opt}$ (the optimal hole concentration) the pressure coefficient dT$_c$/dp should be an anti-symmetric function of $n - n_{opt}$, changing sign exactly at n$_{opt}$. However, experimental data indicate deviations from the perfectly anti-symmetric dependence and additional factors have to be taken into account to understand the pressure effects in HTS. In particular, the suggested negative pressure coefficient of T$_c$ for over-doped materials could not be experimentally confirmed for several compounds [73]. Neumeier and Zimmermann [74] proposed to separate different contributions to the pressure coefficient of T$_c$ and postulated the following relation

$$\frac{dT_c}{dp} = \frac{dT_c^i}{dp} + \frac{\partial T_c}{\partial n}\frac{dn}{dp} \quad . \tag{6}$$

The first term in (6) represents the part of the pressure coefficient that is not due to a change of carrier density while the last term takes account of the pressure-induced charge transfer and its effect on T$_c$. This equation has successfully been used to interpret the pressure coefficients of several HTS compounds such as Y-123, Y-124, and other rare earth-123 superconductors [75]. It should be noted that the structure of the 214-LBCO class of HTS compounds is exceptional in that it has no charge reservoir layer near the CuO$_2$ planes. Consequently, pressure cannot transfer charges from a reservoir to the active layer and the second term in (6) does not contribute. Indeed, it was reported that $dT_c/dp$ of LBCO and LSCO (S for Sr) is positive over the whole doping range [50, 76] and it can be considered as a measure of the first term, $dT_c^i/dp$.

Employing the parabolic relation (5) the pressure derivative of T$_c$ can be written in the more general form [77]

$$\frac{dT_c}{dp} = \frac{\partial T_c}{\partial T_{c,\max}}\frac{dT_{c,\max}}{dp} + \frac{\partial T_c}{\partial a}\frac{da}{dp} + \frac{\partial T_c}{\partial n}\frac{dn}{dp} + \frac{\partial T_c}{\partial n_{opt}}\frac{dn_{opt}}{dp} \quad . \tag{7}$$

The four terms of equation (7) contribute to the total pressure effect on T$_c$ with different magnitude and sign, depending on the actual carrier density (under-doped or over-doped), and $dT_c/dp$ can be a rather complex function. Besides the pressure-induced charge transfer it is not known how n$_{opt}$, T$_{c,\max}$, or the parameter *a* will change with pressure and what the microscopic mechanisms for the assumed changes are.

Changes in the electronic structure as well as Fermi surface instabilities similar to those discussed in Section 2 for LTS compounds could also play an important role in the unusual pressure dependence of T$_c$ of some HTS superconductors. The anomalous doping dependence of dT$_c$/dp in La$_{2-x}$Sr$_x$CuO$_4$ and La$_{2-x}$Ba$_x$CuO$_4$ in conjunction with the anomalous isotope effect reported in both compounds have been interpreted in terms of an abrupt change of the Fermi surface topology (Lifshitz transition) [78]. In the HTS series RSr$_2$Cu$_{2.7}$Mo$_{0.3}$O$_y$ (R rare earth element) an abrupt change of dT$_c$/dp with increasing rare earth ionic radius was reported and a possible electronic transition was proposed [79]. Van Hove singularities (large peaks of the electronic density of states) are common in low-dimensional electronic structures and can cause abnormal behavior upon doping or application of pressure whenever the Fermi energy is close to the singularity. The existence of Van Hove singularities and a pressure-induced change of the Fermi surface topology were proposed to explain the unusual change of slope of $dT_c/dp$ observed in oxygenated $HgBa_2CaCu_2O_{6+\delta}$ [80]. Deviations from the parabolic $T_c(n)$ dependence (5) as, for example, reported in $HgBa_2CuO_{4+\delta}$ [81] will also lead to a more complex pressure dependence of $T_c$ than predicted by the charge transfer model. High-pressure investigations of $HgBa_2CuO_{4+\delta}$ over a wide range of $\delta$ show indeed a change from linear to non-linear $T_c(p)$ behavior with increasing $\delta$ that was ascribed to a possible pressure dependence of the optimal hole density $n_{opt}$ [77].

Another factor that needs to be considered in understanding how pressure affects high-temperature superconductivity is the possible change of the oxygen arrangement in the HTS structure induced by external pressure. The oxygen content in most HTS compounds deviates from the ideal stoichiometric composition contributing to the doping of carriers. For example, in $YBa_2Cu_3O_{7-\delta}$ not all of the possible lattice positions in the CuO-chains are occupied by oxygen and the mobility of oxygen ions along the chains at room temperature is substantial. In the oxygen-doped $La_2CuO_{4+\delta}$ the excess oxygen occupies interstitial positions and can easily be moved as a function of temperature or pressure. This leads to unusual but interesting effects such as macroscopic chemical phase separation in very under-doped $La_2CuO_{4+\delta}$ ($\delta<0.05$) [82]. The re-arrangement of mobile oxygen in HTS induced by pressure will be discussed in more detail in the following Section.

### 3.3  Pressure-Induced Redistribution of Mobile Oxygen in HTS

It is a peculiarity of HTS cuprates that the superconducting state emerges from the insulating state of the "parent" compound by proper doping of charge carriers. This can be achieved by substitution of cations with different valence and by changing the oxygen content of the compound. The latter doping process creates either excess oxygen occupying interstitial sites as in the lattice of $La_2CuO_{4+\delta}$ or oxygen vacancies as, for example, in the CuO chains of $YBa_2Cu_3O_{7-\delta}$. Due to its strong electrostatic potential the mobile oxygen (interstitial or along the chains) will interact with the hole distribution in the $CuO_2$ planes and, if present, with cationic dopants of different valence. In fact, relaxation effects in HTS have been observed as early as 1990. Veal et al. [83] reported a strong dependence of $T_c$ of high-temperature quenched single crystals and ceramic samples of $YBa_2Cu_3O_{7-\delta}$ upon annealing at room temperature resulting in a gradual increase of the superconducting transition temperature by up to 15 K. The annealing effect on $T_c$ was explained by oxygen-vacancy ordering at ambient temperature in the chain region of the structure. Based on a chemical valence model Veal and Paulikas showed a correlation between the oxygen-vacancy distribution and the carrier density in the active $CuO_2$ layer explaining some peculiarities in the $T_c$ versus $\delta$ phase diagram of $YBa_2Cu_3O_{7-\delta}$ [84].

It appears natural that external pressure can have an effect on the oxygen distribution and, consequently, on the properties of the superconductivity and the value of $T_c$ in HTS compounds. This adds additional complexity to the chapter of high-pressure effects in cuprates that was not discussed in detail in the previous sections. The first indication for pressure-induced redistribution of oxygen ions was found in 1991 in the over-doped $Tl_2Ba_2CuO_{6+\delta}$ [56]. In recent years similar effects have been observed in quite different HTS cuprates (under-doped as well as over-doped) and it is suspected that this phenomenon is more or less universal for high-temperature superconductors with mobile oxygen ions.

$Tl_2Ba_2CuO_{6+\delta}$ is a typical over-doped high-$T_c$ cuprate with a maximum $T_c$ close to 90 K at $\delta \approx 0.1$ and zero $T_c$ at $\delta \approx 0.22$ [85]. An unusually large negative pressure coefficient, $d\ln T_c / dp = -0.183\, GPa^{-1}$, was reported for this compound close to $\delta=0.18$ in a piston-cylinder high-pressure clamp (pressure is applied at room temperature) [86]. However, Sieburger and Schilling found striking differences in the pressure dependence of $T_c$ of $Tl_2Ba_2CuO_{6+\delta}$ depending on the temperature at which pressure was applied. Using a He-gas pressure system the large negative pressure effect on $T_c$ of Ref. [86] was confirmed when pressure was increased at room temperature. In contrast, the value of $d\ln T_c / dp$ was positive and much smaller in its magnitude (of the order of 0.01 $GPa^{-1}$) when pressure was imposed below 100 K [56], as shown in Fig. 10. This led to the conclusion that pressure-induced diffusion of mobile interstitial oxygen plays a major role at high temperatures and that application of pressure at low enough temperatures does not change the oxygen configuration since any ionic diffusion process is inhibited at low T.

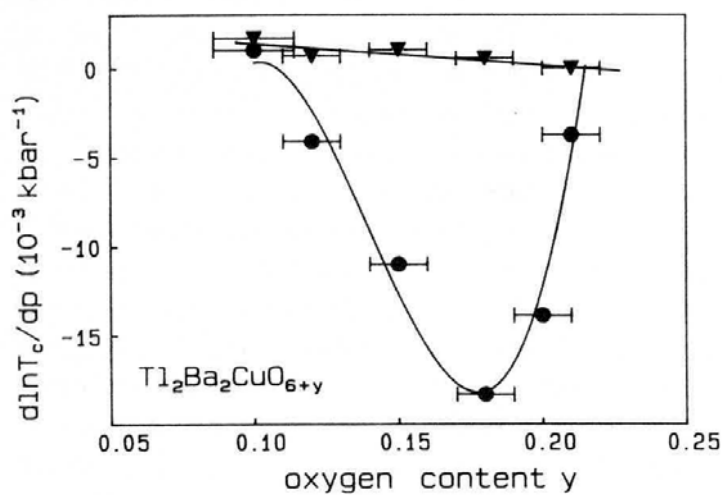

**Fig. 10**: The relative pressure coefficient of $Tl_2Ba_2CuO_{6+\delta}$. Circles: Pressure applied at room temperature. Triangles: Pressure applied at low temperature. (Figure reproduced from Ref. [56]).

Close to the maximum $T_c$ (optimal doping) the two pressure coefficients (for pressure applied at high or low T) are almost identical so that the pressure-induced redistribution of oxygen does not seem to be essential close to the optimal doping level. This result was confirmed by Schirber et al. [87] for $Tl_2Ba_2CuO_{6+\delta}$ with $\delta \approx 0$ ($T_c$=92 K). In the $\delta \to 0$ limit of doping there is no interstitial oxygen in the structure that could be re-arranged by high pressure. Consequently, the value of $d\ln T_c / dp = 0.022 \, GPa^{-1}$ does not depend on the temperature at which pressure was impressed.

The kinetic process of pressure-induced oxygen diffusion and the typical relaxation behavior of $T_c$ were investigated in detail for $Tl_2Ba_2CuO_{6+\delta}$ [88]. A typical $T_c$-relaxation experiment is described in Fig. 11. Two separate relaxation processes have been identified, one low-temperature (LT) relaxation that appears to strongly depend on the concentration of interstitial oxygen $\delta$ and a high temperature (HT) relaxation process the magnitude of which (1 to 2 K change of $T_c$) was found to be independent of $\delta$ [88]. Two characteristic activation energies $E_A$ of 0.25 eV and 0.72 eV have been estimated for LT and HT relaxations, respectively. The value for the HT process is comparable with the activation energy $E_A$=0.96 eV derived from the $T_c$ relaxation in the quenching and annealing experiments of $YBa_2Cu_3O_{7-\delta}$ [83]. $E_A$ for the LT relaxation was shown to be independent of pressure whereas $E_A$ for the HT relaxation of $T_c$ was found to increase with p [89]. The mobile oxygen ions in $Tl_2Ba_2CuO_{6+\delta}$ are interstitial ions residing in bi-layers of $Tl_2O_2$. A hard-sphere model for the diffusion of interstitial oxygen along different diffusion paths was proposed in Ref. [89] and the authors come to the conclusion that the LT relaxation of $T_c$ in $Tl_2Ba_2CuO_{6+\delta}$ is caused by an intra-unit cell migration of oxygen between two equivalent sites with a low activation energy and a vanishing diffusion volume. In contrast, the HT relaxation was ascribed to the diffusion of oxygen ions between different unit cells that requires a partial expansion of the lattice along the diffusion path and is, therefore, characterized by the higher activation energy and a finite diffusion volume.

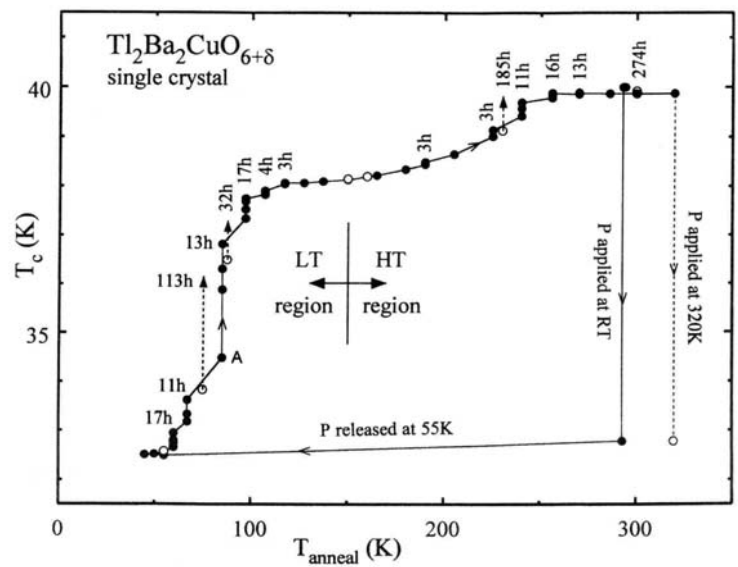

**Fig. 11**: Annealing effect on $T_c$ of a $Tl_2Ba_2CuO_{6+\delta}$ single crystal ($T_c$=40 K). Hydrostatic pressure of 0.6 GPa was first applied at room temperature and then released at 55 K. Subsequent annealing at increasing temperatures for times indicated in the figure results in an increase of $T_c$ towards its ambient value. Two distinct relaxation steps can be distinguished. (Figure reproduced from Ref. [88]).

$T_c$ relaxations under external pressure and subsequent annealing have also been observed in the over-doped $TlSr_2CaCu_2O_{7-\delta}$ (Tl-1212, $T_{c,max}\approx 80$ K) [88]. $T_c$ decreases with increasing oxygen content. The pressure coefficient of $T_c$ when pressure is applied at room temperature significantly depends on the doping state and changes sign from strongly negative close to optimal doping (high $T_c$) to a relatively small, but positive value at higher doping level (lower $T_c$). However, when pressure is changed at low temperature $d\ln T_c/dp$ appeared to be positive over the whole doping range [90]. Relaxation effects of $T_c$ have been observed only at annealing temperatures exceeding 210 K.

$YBa_2Cu_3O_{7-\delta}$ was the first HTS compound with $T_c$ above the temperature of liquid nitrogen [17]. A large amount of scientific work was devoted to this particular compound. In it's optimal doping state ($\delta\approx 0.05$, $T_c=93$ K) the application of pressure had a negligible effect on $T_c$ [48], a property that is common for many of the optimally doped HTS compounds. With increasing $\delta$ $YBa_2Cu_3O_{7-\delta}$ becomes under-doped and more oxygen vacancies are created in the CuO chains. A giant pressure coefficient $dT_c/dp$ of up to 30 K/GPa was reported for very under-doped $YBa_2Cu_3O_{7-\delta}$ [91] (see Fig. 12) when pressure was imposed at room temperature. However, application of pressure below 100 K resulted in a moderate increase of $T_c$ (2 to 4 K/GPa) even for the lowest doping levels. The pressure coefficients of $T_c$ of $YBa_2Cu_3O_{7-\delta}$ are plotted for the whole doping range in Fig. 12 and the enhancement of $dT_c/dp$ for room temperature pressure application is very obvious. This enhancement was attributed to the pressure-induced ordering of the chain oxygen in analogy to the results of the annealing experiments of Veal et al. [83].

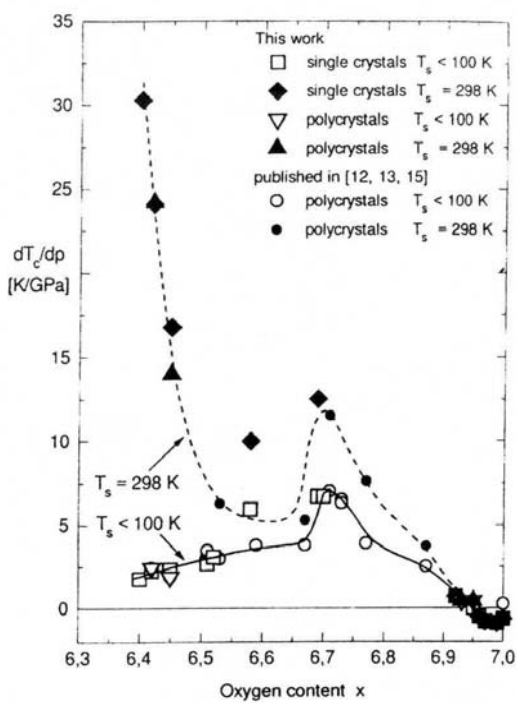

**Fig. 12**: The pressure coefficient of $T_c$ of $YBa_2Cu_3O_{7-\delta}$. Solid curve (open symbols): Pressure applied at low temperature. Broken line (filled symbols): Pressure applied at high temperature. (Figure reproduced from Ref. [91]).

The pressure-induced redistribution of oxygen was investigated in more detail with respect to the characteristic time relaxation of $T_c$ at different temperatures [91,92]. Changing pressure abruptly at room temperature and measuring $T_c$ at different times after storing the pressurized sample at 298 K a typical relaxation behavior as shown in Fig. 13 was observed. The $T_c$ relaxation was suppressed for storage temperatures below 240 K [92]. The time dependence of $T_c$ after each change of pressure is well described by a stretched exponential function, as shown for data on $YBa_2Cu_3O_{6.41}$ [93]. The characteristic relaxation time, $\tau$, increases dramatically with pressure and it was suggested that the $T_c$ relaxation is due to the thermally activated diffusion of $O^{2-}$ ions with $\tau(p,T) = \tau_0 \exp\{E_A(p)/k_B T\}$. The ambient-pressure value of $E_A \approx 0.97$ eV is in excellent agreement with the data of Fietz et al. [91], Veal et al. [83] and with the results of $^{18}O$ tracer diffusion experiments [94]. From the pressure dependence of $E_A$ the activation volume for diffusion could be calculated and was found to be close to the molar volume of $O^{2-}$ [93,95].

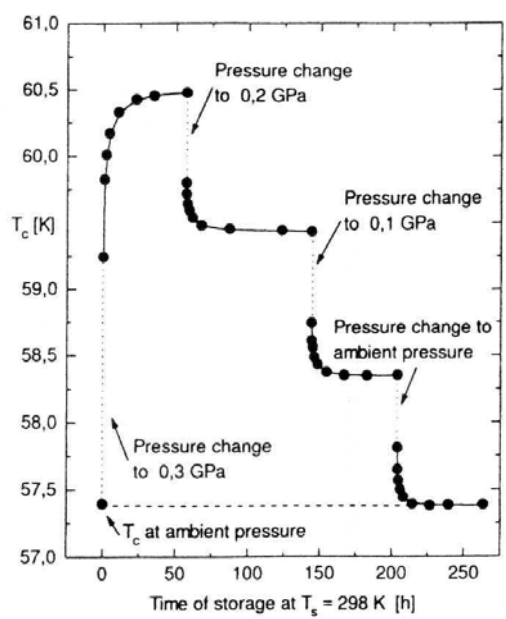

**Fig. 13**: $T_c$ relaxation effects of an under-doped $YBa_2Cu_3O_{6.58}$ single crystal. The sample was annealed at 298 K after pressure changes as indicated. (Figure reproduced from Ref. [91]).

The observed changes of $T_c$ in YBCO with the re-distribution of the chain oxygen was explained by a charge transfer to the active $CuO_2$ layers that depends sensitively on the specific oxygen configuration [83,84]. This charge transfer also affects the electrical transport properties in the normal state. The time-relaxation of the electrical resistivity of $YBa_2Cu_3O_{7-\delta}$ observed after hydrostatic pressure was applied (or released) at room temperature followed exactly the same stretched exponential relaxation function as the corresponding data for $T_c$ with very similar relaxation times, activation energies, and migration volumes [95]. This obvious correlation between the relaxation of $T_c$ and that of the normal-state resistivity provides evidence for the common origin of both effects.

The ordering of oxygen vacancies at ambient pressure was also observed near optimal doping in YBa$_2$Cu$_3$O$_{6.95}$, most notably in thermal expansion measurements [96]. Upon cooling from high temperature a glass-like transition with distinct anomalies in the thermal expansion coefficients around 280 K indicates the rearrangement of the oxygen ions. The largest anomaly is seen in the expansivity along the orthorhombic b-axis. The kinetics of this glass transition is determined by the energy barrier of E$_A$=0.98 eV [96], in agreement with other oxygen ordering studies discussed above. Under hydrostatic pressure the glass transition temperature increased at a rate of 25 K/GPa [97] supporting the assumption that pressure indeed facilitates the rearrangement of the mobile oxygen in the structure of YBCO. There arises the question why for the optimally doped YBa$_2$Cu$_3$O$_{6.95}$ the pressure coefficient of T$_c$ was found to be independent of the temperature at which pressure was applied (see e.g. Fig. 12). According to the universal phase diagram of HTS the T$_c$(n) inverted parabola exhibits a maximum at the optimal carrier density, n$_{opt}$. Any pressure-induced charge transfer mediated by the rearrangement of the mobile oxygen should therefore have a negligible effect on T$_c$ since $dT_c/dn = 0$ for n=n$_{opt}$. Furthermore, the number of oxygen vacancies in YBa$_2$Cu$_3$O$_{6.95}$ is small (δ=0.05) and the change of carrier density in the active layer can be very small at optimal doping although the thermal expansivities show clear anomalies [96]. Therefore, no relaxation effects of T$_c$ or resistivity could be observed in the optimally doped YBa$_2$Cu$_3$O$_{6.95}$.

The study of pressure-induced oxygen diffusion in HgBa$_2$CuO$_{4+\delta}$ is of particular interest since this compound is stable in the whole doping range. The T$_c$ of this compound is a smooth parabolic function of δ covering the under-doped as well as the over-doped region of the phase diagram with a maximum T$_{c,max}$=97 K at δ≈0.22 [81]. $dT_c/dp$ was found to be positive and of the order of 2 K/GPa as long as δ≤0.22 (under-doped to optimally doped) [77]. In the over-doped range (δ>0.22) the shift of T$_c$ was relatively small and a non-linear function of pressure. In the experiments of Ref. [77] pressure was imposed at room temperature. Pressure-induced relaxation processes of T$_c$ have also been observed in over-doped HgBa$_2$CuO$_{4+\delta}$ with typical activation energies of about 0.9 eV (comparable with values of other HTS compounds) and activation volumes of 11 cm$^3$/mol [98].

The La-214 HTS system allows for the simultaneous doping with cations (e.g. Sr$^{2+}$) and anions (O$^{2-}$) and its chemical formula is written in the general form La$_{2-x}$(Sr/Ba)$_x$CuO$_{4+\delta}$. It is important to emphasize that there is a fundamental difference between these two doping options. The cationic dopants (Sr$^{2+}$ or Ba$^{2+}$) replace the La$^{3+}$ ions and reside in lattice site positions. Therefore, they are tightly bounded, immobile at ambient temperature, and randomly distributed among the La positions in the structure. We will use the notation "hard doping" because the dopants are literally frozen at ambient or even elevated temperatures. In contrast, the anionic doping by excess oxygen (notated as "soft doping" hereafter) creates interstitial O$^{2-}$ ions that are highly mobile and will adjust to the change of thermodynamic parameters such as temperature or pressure. In a quantum statistical sense the two types of doping are well distinguished by the way thermodynamic and configuration averaging is conducted. Since doped systems are in general "disordered" systems and many different configurations contribute to the partition function the configuration average has to be performed in addition to the usual thermal average. As pointed out by Brout [99] for the case of hard doping the configuration average has to be taken after the thermal average and it is applied to the thermodynamic potential (free energy) that was calculated for all different dopant configurations. In the case of soft doping, however, the sequence of averaging is the opposite, i.e. the configuration average is applied to the partition function before the thermodynamic averaging.

In $La_{2-x}Sr_xCuO_{4+\delta}$ the total hole density is determined by two parameters, x and $\delta$, and the effects of hard and soft doping can be revealed by controlling both parameters independently. The crucial parameter is $\delta$, the amount of soft excess oxygen doping, that gives rise to a number of interesting physical phenomena not observable in exclusively hard-doped compounds. Structural transitions and macroscopic phase separation have been observed in the super-oxygenated $La_2CuO_{4+\delta}$ [100-107]. For $\delta>0.05$ the superconducting transition proceeds in two well-separated steps with two $T_c$'s at 15 K and 30 K, respectively, suggesting the possibility of electronic phase separation [100]. Interestingly, the two subsequent transitions appear in $La_2CuO_{4+\delta}$ and in $La_{1.985}Sr_{0.015}CuO_{4+\delta}$ above the same critical hole density of $n_h\approx0.085$, as shown in Fig. 14. Care has to be taken in estimating the average carrier density in the $CuO_2$ layers. Similar to the reduced doping efficiency of oxygen observed in $HgBa_2CuO_{4+\delta}$ [81] it was shown that the doping efficiency of oxygen in $La_{2-x}Sr_xCuO_{4+\delta}$ is reduced to 1.3 holes per oxygen when the total hole density exceeds the value of 0.06 holes per Cu [108]. Oxygen-doped and oxygen-strontium co-doped La-214 compounds compared at the same total carrier density (but at different $\delta$, depending on the amount of Sr) exhibit very similar superconducting properties, as demonstrated in Fig. 14 [109]. One superconducting transition is observed with $T_c\approx30$ K in the low-doping range, $n_h<0.085$. At higher doping ($n_h>0.085$) two transitions at 15 and 30 K, respectively, are observed in the magnetic susceptibility. This was interpreted as a signature of electronic phase separation into two superconducting (hole-rich and hole-poor) states that has been predicted to be a general property of two-dimensional hole-like systems [110,111].

The effect of hydrostatic pressure on the superconducting phases in the one- as well as two-transition regions of the phase diagram (Fig. 14) and possible p-induced oxygen migration effects have been investigated in $La_2CuO_{4+\delta}$ and $La_{1.985}Sr_{0.015}CuO_{4+\delta}$ [109]. Hydrostatic pressure was imposed either at room temperature allowing the redistribution of the interstitial oxygen or below 100 K where oxygen diffusion is kinetically inhibited. An unusually large pressure coefficient, $dT_c/dp=10$ K/GPa, was detected in the low doping range ($n_h<0.085$) for pressures applied at 296 K. In contrast, the p-shift of $T_c$ was much smaller (2 to 3 K/GPa) when pressure was changed at low T. This remarkable difference led to the conclusion that pressure induces a redistribution of the soft dopants (interstitial oxygen), similar to the effects observed in YBCO. The extraordinary large $dT_c/dp$ in $La_{2-x}Sr_xCuO_{4+\delta}$, however, cannot be explained by a charge transfer from a charge reservoir to the active $CuO_2$ layers (as in YBCO) since there is no charge reservoir in the 214-structure. Hall effect measurements have confirmed that the average density of free carriers does not change with the application of pressure [112,113]. It was therefore suggested that the pressure facilitates a redistribution of charges resulting in a local enhancement of the carrier density and the separation into hole-rich and hole-poor phases [109]. This charge separation is made possible by the mobile oxygen ions that adjust their distribution to provide the needed screening and to compensate the Coulomb repulsion between the holes in the $CuO_2$ layers. When the oxygen ions are frozen in their positions and p is changed at low T the $O^{2-}$ act like pinning centers for the holes and the p-induced charge separation is largely suppressed. This explains the smaller $dT_c/dp$ for low-temperature application of pressure.

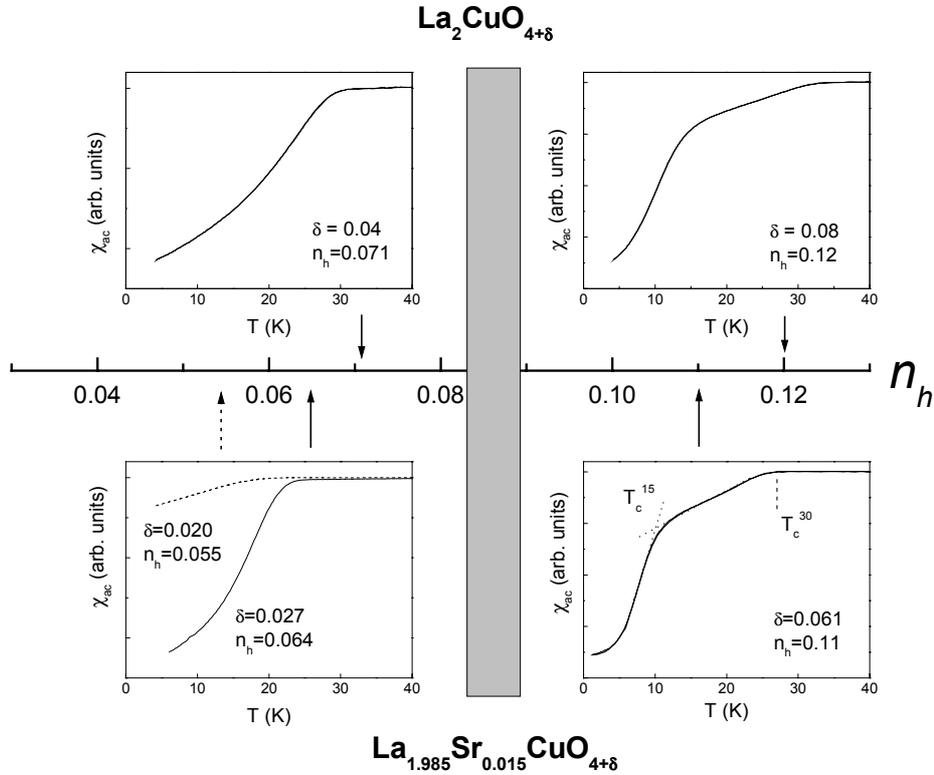

**Fig. 14**: Typical ac susceptibility curves in the phase diagram of $La_2CuO_{4+\delta}$ and $La_{1.985}Sr_{0.015}O_{4+\delta}$ as function of the average hole concentration, $n_h$. Two superconducting transitions are observed for $n_h>0.085$. (Figure reproduced from Ref. [109]).

The pressure effects on the 15 K and 30 K superconducting transitions for higher doping levels ($n_h>0.085$) are opposite in sign ($T_c^{(15)}$ decreases and $T_c^{(30)}$ increases with p) when p is applied at 296 K. Changing pressure at low T (<100 K), however, barely affects the 15 K transition temperature but it increases $T_c^{(30)}$ at a moderate rate of about 3 K/GPa. These differences in the pressure effects on $T_c^{(15)}$ and $T_c^{(30)}$ and their dependence on the temperature at which pressure is changed show the important role the mobile oxygen dopants play in $La_{2-x}Sr_xCuO_{4+\delta}$ in stabilizing superconductivity. The results have been interpreted in terms of pressure-induced electronic phase separations. Further details can be found in Ref. [109].

The fundamental difference of cationic (hard) and anionic (soft) doping becomes drastically visible in the high-pressure investigation of very different HTS compounds as discussed above. Wherever doping is achieved by introducing mobile oxygen ions (or vacancies) the application of pressure at room temperature results in a redistribution of the $O^{2-}$ ions and an increased response of the superconducting state (e.g. enhancement of $dT_c/dp$). The thermodynamically stable state under imposed pressure is characterized by a different oxygen configuration than the ambient-pressure state. This has to be taken into account as an additional dimension in the understanding of the pressure effects in HTS. When pressure is applied at low temperature and oxygen diffusion is inhibited the resulting high-pressure state is actually away from thermodynamic equilibrium. The resulting pressure effects do not reflect the complex response of the superconductor to external compression that has to include the correlation between the hole carriers and the oxygen ions. The $T_c$-relaxation experiments discussed above therefore indicate the transition from a metastable state to the thermodynamically stable state at a given pressure.

## 4. PRESSURE EFFECTS IN SOME UNCONVENTIONAL SUPERCONDUCTORS

### 4.1 Pressure Effects in MgB$_2$ and Isostructural Intermetallic Compounds

Soon after the discovery of superconductivity in MgB$_2$ at temperatures as high as 39 K [114] one of the most disputed questions was whether or not the superconductivity in this simple intermetallic compound can be understood on the basis of a phonon-mediated BCS-like pairing mechanism. The unusually high $T_c$ was attributed to the strong electron-phonon interaction and the high phonon frequency of lattice vibrations involving mainly the light boron element [115]. Alternatively, Hirsch proposed an explanation in terms of a "universal" theory of hole superconductivity conjecturing that the mechanism of superconductivity in MgB$_2$ is similar to that in cuprate superconductors and that the pairing of heavily dressed holes in almost completely filled bands is driven by a gain in kinetic energy. An increase of $T_c$ with pressure was predicted if pressure does reduce the in-plane boron-boron distance [116]. Evidence for hole-type carriers was indeed derived from early measurements of the thermoelectric power [117]. The pressure coefficient of $T_c$ of MgB$_2$ was first measured in a piston-cylinder clamp with liquid pressure transmitting medium. The results are shown in Fig. 15.

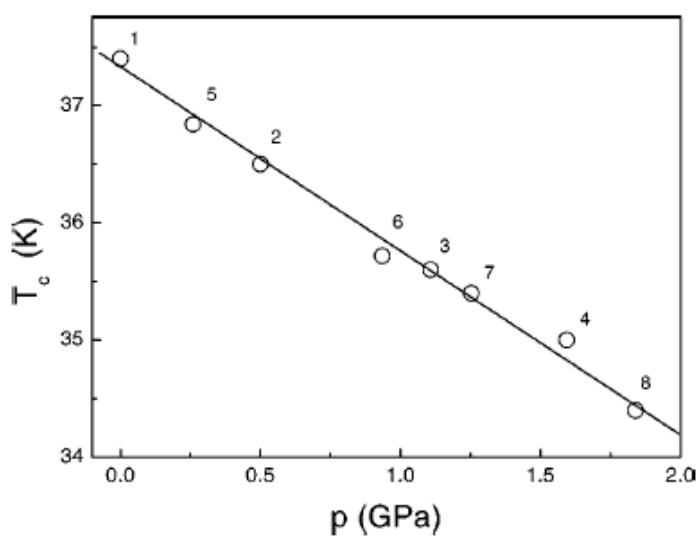

**Fig. 15**: Effect of hydrostatic pressure on the superconducting transition temperature of MgB$_2$. The numbers next to the data points indicate the sequence of pressure changes. (Figure reproduced from Ref. [117]).

The pressure coefficient of $T_c$ was found to be negative and of the order of -1.6 K/GPa [117] in very good agreement with calculations of the elastic and electronic properties of $MgB_2$ [118]. Therefore, the theory of hole superconductivity appeared to be less favorable to explain the mechanism of superconductivity in $MgB_2$. Since the pressure coefficient of $T_c$ is essential to discriminate between the different proposed models for $MgB_2$, a large number of follow-up high-pressure experiments have been conducted. In all experiments employing hydrostatic pressure $dT_c/dp$ was found to be negative but its value scattered between $-1$ and $-2$ K/GPa in most investigations. The different values of $dT_c/dp$ reported by various groups have been a matter of a separate discussion (see for example Ref. [119, 120]). It was suggested that the larger values of $dT_c/dp$ up to $-2$ K/GPa might be related to the sensitivity of the superconducting state of $MgB_2$ to stress introduced by the freezing of the liquid pressure medium upon cooling [121]. However, this suggestion could not be confirmed in alternative experiments [122]. Instead, a correlation between "sample quality" as expressed by small variations of $T_c$ and the magnitude of $dT_c/dp$ was proposed [123]. The role of defects in the structure of $MgB_2$ was discussed in detail and it was found that the increase of the defect density results in an increase of the lattice strain, an increase of the c-axis length, a decrease of the resistivity ratio $\rho(300K)/\rho(T_c)$, and a decrease of $T_c$ [120,124]. The linear increase of $|dT_c/dp|$ with decreasing $T_c$ could therefore be related to the increase of the defect density in "poor" samples [120,123].

The current microscopic understanding of the superconductivity in $MgB_2$ reveals an unusual superconducting gap structure where supercarriers are formed simultaneously in two bands at the Fermi energy and two superconducting gaps of different magnitude open at $T_c$ [125]. This two-gap scenario was confirmed experimentally by a number of investigations including heat capacity measurements [126], Raman scattering [127], and tunneling spectroscopy [128]. To understand the pressure effects in superconducting $MgB_2$ the p-induced changes of the microscopic parameters of the two bands as well as the coupling between the bands need to be taken into account. Experimentally, the measurement of the heat capacity ($C_p$) under high-pressure conditions appears to be of interest since changes in the superconducting gap structure are reflected in the temperature dependence of $C_p$ [126]. No such experiments have been conducted so far.

The excitement about the high superconducting transition temperature of $MgB_2$ initiated the search for similar intermetallic compounds with the same lattice structure. The most prominent materials found superconducting are CaAlSi and SrAlSi as well as the Ga-based series (Ca,Sr,Ba)GaSi [129]. These compounds are isostructural to $MgB_2$ with Al (or Ga) and Si occupying randomly the boron sites in the C32 structure of $MgB_2$. The critical temperatures of these new materials turned out to be moderate with the highest $T_c=8$ K found in CaAlSi. Although CaAlSi and SrAlSi are very similar electronically as well as structurally (the major difference is the 7.5 % larger c-axis lattice parameter of SrAlSi), the application of hydrostatic pressure revealed a fundamental difference of their superconducting states [130]. The superconducting $T_c$ of SrAlSi decreases linearly at a rate of $-0.12$ K/GPa, as shown in Fig. 16. However, $T_c$ of CaAlSi increases with pressure in a nonlinear way with an initial pressure coefficient of 0.21 K/GPa (Fig. 16). Heat capacity experiments (Fig. 17) indicate that superconductivity in SrAlSi is well described by the BCS theory in the weak-coupling limit, however, the data for CaAlSi show the typical signature of a very strong-coupling superconductor [21,130].

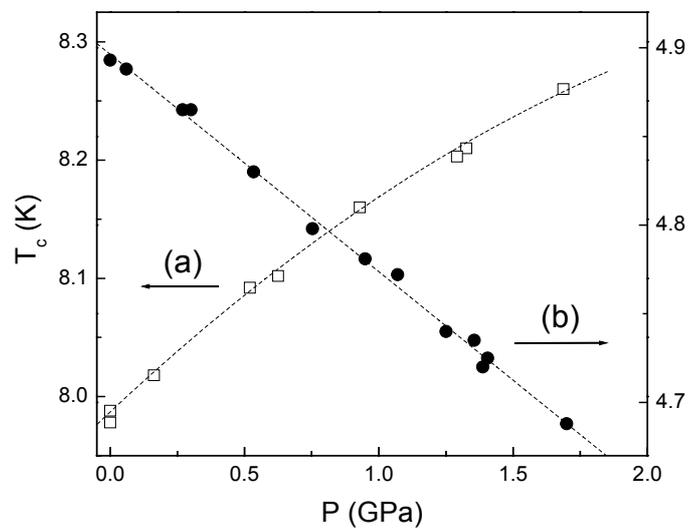

**Fig. 16**: Effect of hydrostatic pressure on the superconducting transition temperatures of (a) CaAlSi (open squares, left scale) and (b) SrAlSi (filled circles, right scale).

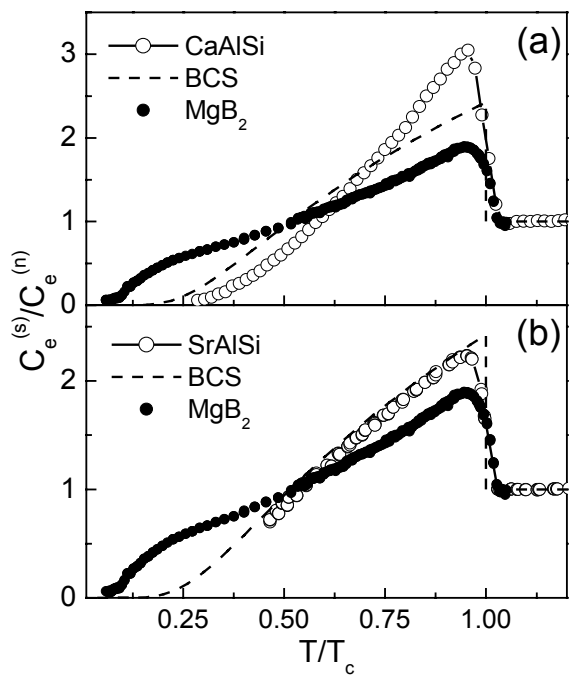

**Fig. 17**: Normalized heat capacity of (a) CaAlSi and (b) SrAlSi. The dashed line shows the BCS function. For comparison the heat capacity of $MgB_2$ is included (filled symbols).

Calulations of the electronic band structure and the electron-phonon coupling could not explain the qualitative disparity in the superconducting properties of the two compounds [131]. It was therefore speculated [131] that a soft mode in CaAlSi could lead to an enhancement of the electron-phonon coupling in CaAlSi and to an unusual structure of the Eliashberg function [21]. The possible existence of a soft mode in CaAlSi might be an indication that the compound is very close to an electronic or structural instability with the dramatic effect on its superconductivity resulting in the "unusual" behavior discussed above. The experimental search for this soft mode and its consideration in a theoretical description of superconductivity in the C32 intermetallic compounds will facilitate our principal understanding of superconductivity in "unconventional" compounds.

## 4.2 Pressure Control of Dimensionality in the New Superconductor $Na_xCoO_2*yH_2O$

The important role of dimensionality in superconductivity has been considered for many decades. With the rising age of HTS cuprates other low-dimensional superconductors have attracted renewed attention. The recent discovery of superconductivity in the sodium-doped cobalt oxyhydrate $Na_xCoO_2*yH_2O$ ($x\approx1/3$, $y\approx4/3$) has raised fundamental questions about the role of low dimensionality, the pairing symmetry, and the unusual properties of this unconventional superconductor [132]. The intercalation of water molecules in between the $CoO_2$ layers forces the c-axis lattice constant to expand by 77 % and increases the anisotropy of the structure in reducing the coupling between the $CoO_2$ layers. It was, therefore, suggested that the superconductivity observed below 5 K arises in the quasi-2d $CoO_2$ layers and analogies to the $CuO_2$ layers of the HTS cuprates have been drawn [132]. Indeed, experimental evidence for a nodal superconducting order parameter and the existence of line nodes in the gap function below $T_c$ suggests a strong similarity of the superconducting state in $Na_xCoO_2*yH_2O$ to the d-wave superconductivity in HTS cuprates [133,134]. With the weak interactions between the $CoO_2$ layers the compressibility along the c-direction is expected to be particularly large. Application of hydrostatic pressure can be used to tune the coupling between the superconducting $CoO_2$ layers and it provides vital insight into the role of dimensionality in stabilizing the superconducting state.

We have found that pressure reduces the critical temperature with an unusual non-linear pressure dependence of $T_c$ [135], as shown in Fig. 18. The negative $dT_c/dp$ of $Na_xCoO_2*yH_2O$ was explained by the larger compression of the c-axis that drives the structure away from the two-dimensional character. Subsequent high-pressure x-ray investigations have confirmed our suggestion and have shown that the compressibility of the c-axis indeed exceeds the a-axis value by a factor of 4.5 [136]. The pressure effects on $T_c$ and the lattice constants provide strong evidence for the two-dimensional character of superconductivity in $Na_xCoO_2*yH_2O$. A further increase of the c-axis should result in higher $T_c$ values and, possibly, new superconducting phenomena could be detected. A very recent structural study [137] of $Na_xCoO_2*yH_2O$ single crystals has shown that the water content can be increased to y=1.8 with an increase of the c-axis lattice parameter from 19.71 Å (y=1.3) to 22.38 Å (y=1.8). However, low-temperature studies and the search for superconductivity in this compound have not yet been conducted.

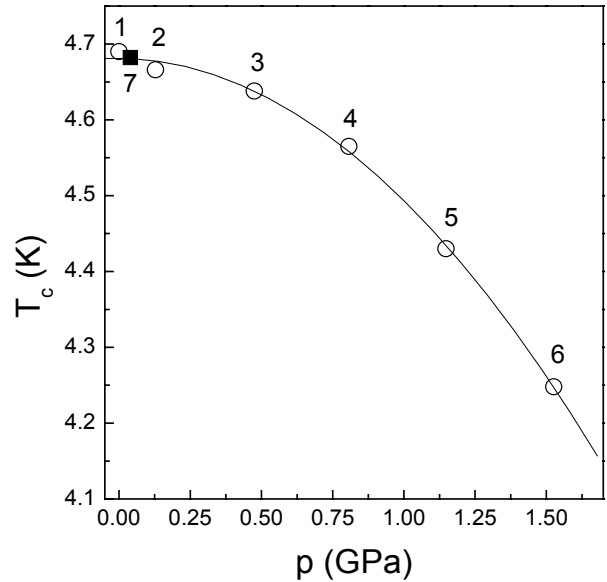

**Fig. 18**: Pressure dependence of $T_c$ of the layered cobalt oxyhydrate $Na_xCoO_2*yH_2O$. The numbers indicate the sequence of pressure changes. The ambient-pressure $T_c$ recovered after complete release of pressure (filled square).

The non-linearity of $T_c(p)$ is interesting by itself and it suggests the possible existence of large dispersion in the energy spectrum near the Fermi surface [135]. The average pressure change of $T_c$ is $d\ln T_c/dp = -0.07\, GPa^{-1}$. Above 1 GPa $T_c$ decreases at a rate of $d\ln T_c/dp = -0.1\, GPa^{-1}$. It is interesting to note that these values compare favorably with the pressure coefficients of electron-doped high-temperature superconductors with a similar $T_c$ [138]. In $Na_xCoO_2*yH_2O$ the charge carriers are electrons. The current high-pressure results may suggest an interesting similarity of the layered $Na_xCoO_2*yH_2O$ and the HTS cuprates.

### 4.3 Competition of Superconducting and Magnetic Orders in Superconducting Ferromagnets as Revealed by High-Pressure Experiments

In the field of high-temperature superconductivity the coexistence of ferromagnetism and superconductivity in a class of Ru-based HTS compounds has raised considerable attention. In these materials (e.g. $RuSr_2GdCu_2O_8$) ferromagnetic order arises at relatively high temperature ($\approx$130 K) and superconductivity sets in at about 45 K. The question if and how these two antagonistic states of matter can coexist has been a focal point of intense discussion (for a short review of the topic see e.g. Ref. [139]). The response of the two states, magnetism and superconductivity, to applied pressure will provide more insight into the way they coexist with one another.

The superconducting ferromagnets are commonly synthesized in polycrystalline form with relatively weak grain-grain connectivity. The weak links across the grain boundaries give rise to unusual effects, for example in the magnetoresistance measurements below $T_c$ [140]. The superconducting transition as observed in resistivity or zero-field cooling magnetization measurements proceeds in two steps, the intra-grain transition is followed by an inter-grain transition (phase coherence across the grain boundaries) at lower temperature. Both transitions can be separated in ac-susceptibility or transport measurements [141,142]. It is essential to focus the high-pressure investigation onto the pressure effects on the bulk superconducting properties and the intra-grain superconducting $T_c$.

The pressure dependences of both, the ferromagnetic transition temperature, $T_m$, and the intra-grain superconducting $T_c$, of $RuSr_2GdCu_2O_8$ were estimated by measuring the ac susceptibility in a piston-cylinder clamp cell [143]. The results for $T_c(p)$ and $T_m(p)$ are shown in Fig. 19. Both transition temperatures increase with pressure, but at different rates. The magnitude of the relative pressure shift of $T_m$, $d\ln T_m / dp = 0.054\, GPa^{-1}$, was found to be about twice as large as $d\ln T_c / dp = 0.025\, GPa^{-1}$, indicating that the stabilizing effects of pressure on the magnetic state are much stronger than on the superconductivity.

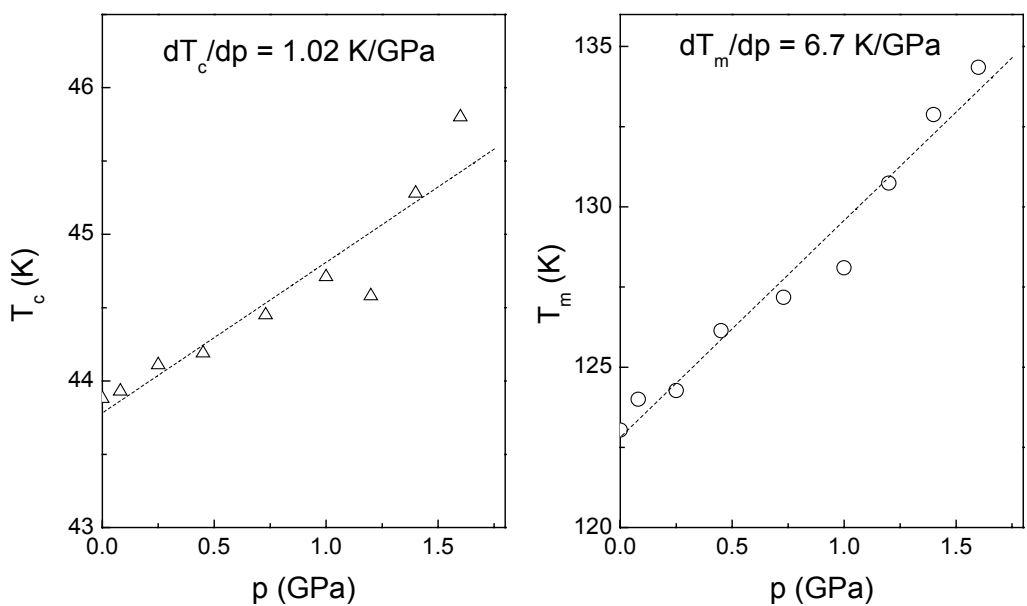

**Fig. 19**: Pressure dependences of the intra-grain $T_c$ (left) and the ferromagnetic $T_m$ (right) of $RuSr_2GdCu_2O_8$.

The absolute pressure coefficient of the intra-grain $T_c$, $dT_c/dp \approx 1\, K/GPa$, is surprisingly small if compared with similar, but non-magnetic high-$T_c$ compounds in a comparable doping state. $RuSr_2GdCu_2O_8$ is known to be a typical under-doped cuprate with a hole density of about 0.06 to 0.07 [141]. The pressure coefficients of $T_c$ of different HTS cuprates at a similar doping level, as for example $La_{2-x}(Sr, Ba)_xCuO_4$, $YBa_2Cu_3O_{7-\delta}$, or $YBa_2Cu_{3-x}M_xO_{7-\delta}$, are all larger by a factor of 3 to 4 than the $dT_c/dp$ of the superconducting ferromagnet $RuSr_2GdCu_2O_8$. This leads us to propose that the magnetic and superconducting states are not completely independent but they compete with each other [143]. Due to this competition the stronger enhancement of the ferromagnetic phase results in a reduced (as compared to similar HTS compounds) pressure effect on $T_c$. Under external pressure the magnetism in $RuSr_2GdCu_2O_8$ is stabilized on the expense of the superconducting state.

## 5. SUMMARY

Since the discovery of superconductivity in 1911 the effects of high pressure on the superconducting state and $T_c$ have been of prominent interest. Pressure was very successful in inducing superconductivity in elements and compounds that are not superconducting at ambient pressure. The total number of elements that turn superconducting at high pressure now almost equals the number of ambient pressure superconducting elements. Especially the high-pressure superconductivity detected in the alkali metal lithium sparks new hope in the ongoing search for superconductivity with an expected unprecedented high $T_c$ in dense metallic hydrogen. The developments of new techniques have pushed the limits of static pressure generation into the range of several hundred GPa and the dream to squeeze hydrogen superconducting may be realized very soon. Large positive pressure coefficients of $T_c$ found in the high-temperature superconducting cuprates have revealed the final clues that led to the synthesis of the first HTS compounds with $T_c$'s above the temperature of liquid nitrogen. The current record $T_c=164$ K was achieved in $HgBa_2Ca_2Cu_3O_{8+\delta}$ under high-pressure conditions.

The physical understanding of the pressure effects on superconductivity is very complex, even for conventional superconductors. Within the phonon-mediated pairing models (BCS in weak-coupling limit or Eliashberg theory for strong coupling) a number of electronic and phononic parameters that depend on pressure with an immediate effect on the transition temperature have to be considered. The equation for $T_c$ derived by McMillen, for example, involves three parameters, the average phonon frequency, the electron-phonon coupling constant, and the screened Coulomb pseudopotential. All three parameters depend on pressure via the phonon energy, the density of states, the electronic matrix elements, the screened Coulomb interaction, etc. in a non-trivial analytical way. A first-principle theory of the pressure effects in superconductivity has to account for the pressure dependence of all parameters of the phonon and electron systems as well as their interactions. In a semi-empirical treatment the pressure dependence of the density of states and the phonon frequency is frequently considered and all other parameters are assumed to be independent of pressure. This simplified procedure was successful in describing the pressure effects on $T_c$ in some simple compounds such as superconducting lead.

Abnormal pressure effects observed in several superconducting elements (Tl, Re) have been explained by electronic instabilities of the Fermi surface (change of topology from closed to open Fermi surface) induced by pressure. High-pressure investigations of superconductivity in some A15 compounds have revealed the superior role of the electronic excitation spectrum with peaks or singularities close to the Fermi energy in stabilizing superconductivity with relatively high $T_c$-values and inducing the phonon softness that leads to structural instabilities close to the superconducting transition.

In high-temperature superconducting cuprates the effect of pressure on $T_c$ strongly depends on the doping state, i.e. the number of holes in the active $CuO_2$ layer. The universal parabolic law that applies to a majority of HTS compounds expresses $T_c$ as an inversed parabolic function of the hole number, n, with an optimal hole density (maximum $T_c$) near 0.16. Because of this $T_c(n)$ dependence $T_c(p)$ is largely determined by a pressure-induced charge transfer from the charge-reservoir block to the $CuO_2$ layer. This charge-transfer model explains the observed sign change of the pressure coefficient of $T_c$ in crossing from the under-doped region (n<0.16) into the over-doped range (n>0.16). However, there are several exceptions in the class of HTS compounds that either do not follow the parabolic $T_c(n)$ dependence in the full range of doping or that show a more complex pressure dependence of $T_c$ contradicting the simple charge transfer model. The existence of Van Hove singularities and the pressure-induced change of the Fermi surface topology have therefore been proposed (e.g. in $HgBa_2CaCu_2O_{6+\delta}$) to account for the abnormal behavior of $T_c(p)$.

An increased complexity of pressure effects have been reported in HTS systems with weakly bound, mobile oxygen ions, such as vacancies in the CuO-chains of Y-123 (YBCO) providing the possibility of oxygen migration along the chains, or interstitial excess oxygen in Tl-2201 and La-214. It is shown that, for several different HTS compounds, the application of pressure results in a change of the distribution of these mobile oxygen ions with a large impact on the pressure dependence of $T_c$. Therefore, the pressure effects depend on the temperature at which pressure is imposed. If this temperature is low enough the oxygen migration is inhibited and the application of pressure drives the HTS system away from thermal equilibrium with a qualitatively different dependence $T_c(p)$. Consequently, $T_c$ relaxation effects have been observed with relaxation times characteristic for diffusion of oxygen ions along specific diffusion paths.

In the superoxygenated La-214 HTS compound the interstitial oxygen facilitates the electronic phase separation into hole-poor and hole-rich states at different doping levels. Pressure applied at ambient temperature increases the tendency to phase separation. The pressure-induced re-arrangement of oxygen ions also results in an enhancement of the pressure coefficient of $T_c$ up to 10 K/GPa. The high-pressure investigations reveal the fundamental difference of cation doping and oxygen doping in HTS compounds and emphasize on the important role of mobile oxygen in stabilizing superconductivity with high $T_c$. Since similar effects have been observed in quite different HTS cuprates it was suggested that the tendency to electronic phase separation and the oxygen ordering phenomena (also induced be pressure) are correlated and represent a universal feature of HTS.

Several examples are discussed where high-pressure investigations have led to a deeper understanding of the underlying physics of specific superconductors or have raised new questions about it. Among them are the pressure effects on superconductivity in $MgB_2$ and the isostructural intermetallic compounds CaAlSi and SrAlSi. In the former case the negative pressure coefficient of $T_c$ provided strong support for the phonon-mediated BCS-like pairing mechanism in contrast to alternative theories. For the latter two compounds the opposite pressure coefficients of $T_c$ in conjunction with large differences in the electron-phonon coupling strengths derived from heat capacity measurements led to speculations about the possible existence of a soft phonon in CaAlSi driving the system close to a structural instability. The suggestion is still awaiting experimental or theoretical verification.

In the sodium-doped cobalt oxyhydrate, $Na_xCoO_2*yH_2O$, the decrease of $T_c$ with hydrostatic pressure was understood in terms of an increased coupling between the $CoO_2$ layers due to a larger compressibility of the c-axis as compared to the a-axis. The compression reduces the 2d character of the water-intercalated structure and the negative pressure coefficient of $T_c$ lends indirect support to the assumption that the high anisotropy (or the low-dimensionality) of the structure is crucial for the understanding of the superconductivity in the compound.

Pressure effects on the ferromagnetic and superconducting transitions of the HTS superconducting ferromagnet $RuSr_2GdCu_2O_8$ are discussed and lead to the conclusion that both antagonistic states of matter compete with one another in the compound. The imposed pressure favors the magnetic state on the expense of the superconducting state that leads to a smaller than expected pressure coefficient of the superconducting $T_c$.

**Acknowledgements**

This work is supported in part by NSF Grant No. DMR-9804325, the T.L.L. Temple Foundation, the John J. and Rebecca Moores Endowment, and the State of Texas through the Texas Center for Superconductivity at the University of Houston and at Lawrence Berkeley Laboratory by the Director, Office of Energy Research, Office of Basic Energy Sciences, Division of Materials Sciences of the U.S. Department of Energy under Contract No. DE-AC03-76SF00098.